\documentclass{article}

\usepackage{arxiv}

\usepackage[utf8]{inputenc} % allow utf-8 input
\usepackage[T1]{fontenc}    % use 8-bit T1 fonts
\usepackage{url}            % simple URL typesetting
\usepackage{booktabs}       % professional-quality tables
\usepackage{amsfonts}       % blackboard math symbols
\usepackage{nicefrac}       % compact symbols for 1/2, etc.
\usepackage{microtype}      % microtypography
\usepackage{lipsum}		% Can be removed after putting your text content
\usepackage{graphicx}
\usepackage{lineno}
\usepackage[utf8]{inputenc}
\usepackage{lipsum}
\usepackage{multicol}
\usepackage[table]{xcolor}
\usepackage{tikz}
\usepackage{amsmath}
\usepackage{amsthm}
\usepackage{flushend}
\RequirePackage{doi}

\definecolor{bar}{rgb}{0.557,0.663,0.859}
\definecolor{back}{rgb}{0.851,0.882,0.949}
\newcommand{\DrawPercentageBar}[1]{%
	\begin{tikzpicture}
	\fill[color=black]   (0.0 , 0.0) rectangle (#1*3ex , 1.5ex );
	\fill[color=lightgray] (#1*3ex  , 0.0) rectangle (3.0ex, 1.5ex);
	\end{tikzpicture}%
}

\title{Securing Microservices and Microservice Architectures: A Systematic Mapping Study}

\author{ {\hspace{1mm}Abdelhakim Hannousse}%\thanks{Use footnote for providing further
	\\
	Department of Computer Science\\
	Universté 8 Mai 1945, Guelma\\
	 BP 401, Guelma 24000, Algeria\\
	\texttt{hannousse.abdelhakim@univ-guelma.dz} \\
	\And
	{\hspace{1mm}Salima Yahiouche} \\
	Department of Computer Science\\
	LRS laboratory, Badji Mokhtar University\\
	BP 12, Annaba 23000, Algeria\\
	\texttt{yahiouche.salima@univ-annaba.dz} \\
}

\begin{document}
\maketitle

\begin{abstract}
\textbf{Context.} Microservice architectures (MSA) are becoming trending alternatives to existing software development paradigms notably for developing complex and distributed applications. Microservices emerged as an architectural design pattern aiming to address the scalability and ease the maintenance of online services. However, security breaches have increased threatening availability, integrity and confidentiality of microservice-based systems.\\
\textbf{Objective.} A growing body of literature is found addressing security threats and security mechanisms to individual microservices and microservice architectures. The aim of this study is to provide a helpful guide to developers about already recognized threats on microservices and how they can be detected, mitigated or prevented; we also aim to identify potential research gaps on securing MSA.\\
\textbf{Method.} In this paper, we conduct a systematic mapping in order to categorize threats on MSA with their security proposals. Therefore, we extracted threats and details of proposed solutions reported in selected studies. Obtained results are used to design a lightweight ontology for security patterns of MSA. The ontology can be queried to identify source of threats, security mechanisms used to prevent each threat, applicability layer and validation techniques used for each mechanism.\\
\textbf{Results.} The systematic search yielded 1067 studies of which 46 are selected as primary studies. The results of the mapping revealed an unbalanced research focus in favor of external attacks; auditing and enforcing access control are the most investigated techniques compared with prevention and mitigation. Additionally, we found that most proposed solutions are soft-infrastructure applicable layer compared with other layers such as communication and deployment. We also found that performance analysis and case studies are the most used validation techniques of security proposals.\\
\textbf{Conclusion.} More researches are needed for securing MSA properly. We advocate more researches addressing internal attacks, proposing mitigation techniques and concerning more layers of MAS including communication and deployment.
\end{abstract}

% keywords can be removed

\section{Introduction}
Nowadays, systems are becoming more complex, larger and more expensive due to the rapid growth of requirements and adoption of new technologies. Moreover, due to competitors, many companies need to make changes to their systems as fast as possible and without affecting their systems availability. This requires appropriate designs, architectural styles and development processes. Software engineering provides different paradigms to partially meet those needs by decomposing software systems into fine-grained software units for better modularity, maintainability and reusability, and hence reduce time-to-market. 

Recently, microservice architectures (MSA)~\cite{Yarygina2018a} has emerged as a new architectural style allowing building software systems by composing lightweight services that perform very cohesive business functions. Microservices are the mainstay of MSA. A mircoservice is a fine-grained software unit that can be created, initialized, duplicated, and destroyed independently from other microservices of the same system. Moreover, micorservices can be deployed across heterogeneous execution platforms over the network. Using microservices enables high scalability and flexibility of large scale and distributed software systems. 

Although the advantages brought by adopting microservice architectures in developing complex systems, MSA as a novel technology comes with many flaws~\cite{Baskarada2018} and security is one of the serious challenges that need to be tackled~\cite{Yarygina2018a}. In fact, security is a longstanding problem in networking systems, but with microsevices, security becomes more challenging. This is due to the large number of entry points and overload on communication traffic emerged by decomposing systems into smaller, independent and distributed software units. Moreover, trusts cannot simply be established between individual microservices in the network that often come from different and unknown providers. 

Due to the massive attacks reported on companies adopting MSA such as Netflix and Amazon\footnote{https://threatpost.com}, dealing with security breaches became an urgent need. Several works in the literature have noticed the need to investigate security of MSA~\cite{Yarygina2018a,Dragoni2017,Alshuqayran2016}. However, security threats are diverse and are continually increasing. Security proposals are also increasing and varies from securing individual microservices into complete architectures and infrastructures.

In this article, we conduct a systematic mapping study to uncover the main threats menacing the security of microservice-based systems. We systematically identify existing studies addressing threats and proposing security solutions to MSA. We apply a thorough protocol to extract, classify, and organize reported threats with the security solutions proposed to mitigate and prevent them. The contributions of the study can be summarized as: 

\begin{enumerate}
	\item identify the most relevant threats concerning microservices and microservice architectures
	\item point out the set of security mechanisms used to detect, mitigate and prevent those threats
	\item determine the set of techniques and tools used to examine and validate proposed solutions
	\item end up with a lightweight ontology for security in microservice architectures
\end{enumerate}

The reminder of this paper is structured as follows: section~\ref{sec:bg} gives a succinct background on used techniques and approaches in this paper, specifically, microservice architectures and systematic mapping elaboration process; section~\ref{sec:rw} overviews and discusses related works; section~\ref{sec:sms} details our research methodology; section~\ref{sec:results} presents and discusses the mapping results; section~\ref{sec:ontology} presents the proposed ontology for MSA; section~\ref{sec:validity} discusses threats to validity related to the study and section~\ref{sec:conclusion} concludes the paper.

\section{Background}
\label{sec:bg}
\subsection{Micorservice Architectures}
\begin{figure}[h]
	\centering
	\includegraphics[width=.5\textwidth]{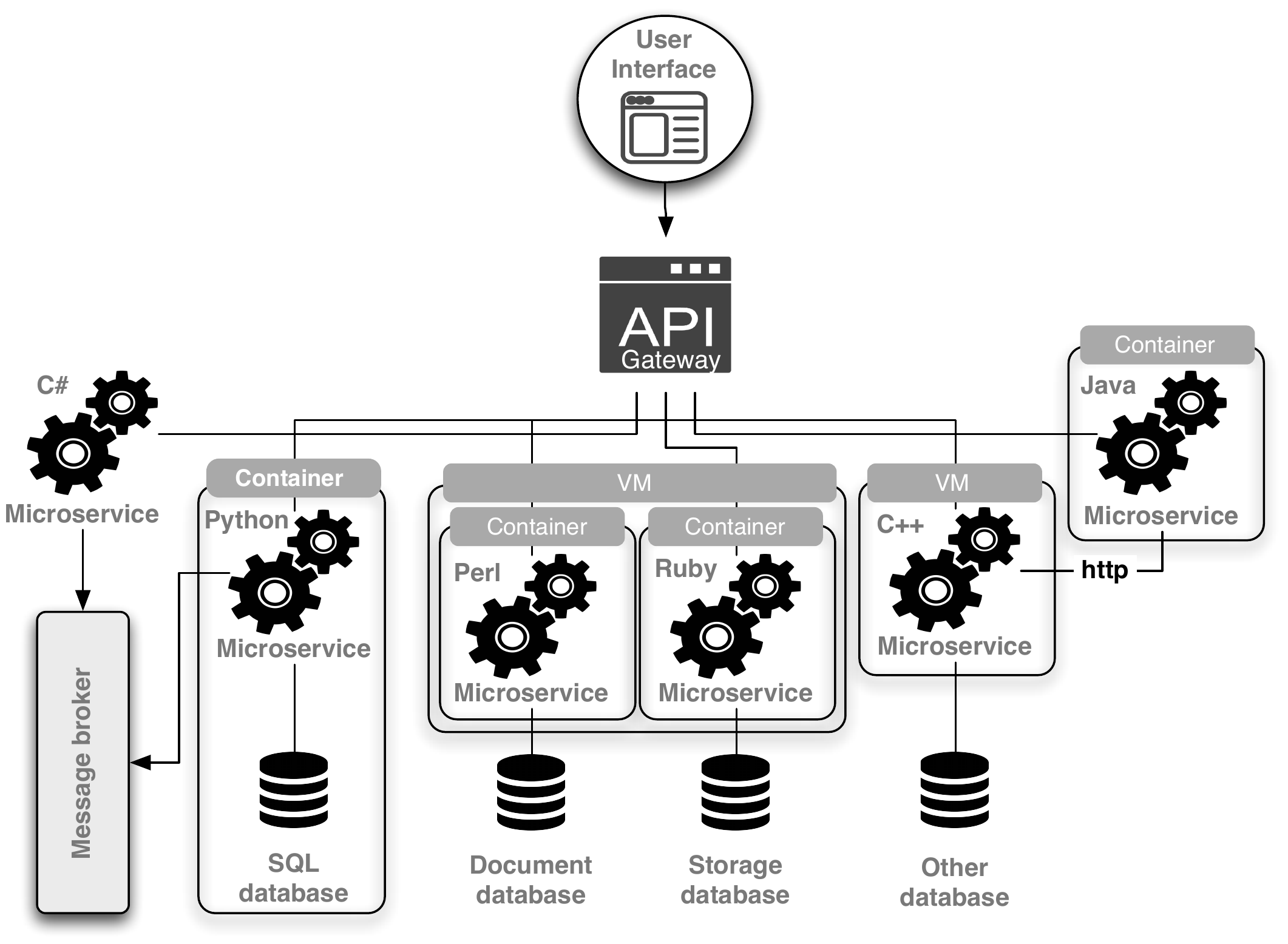}
	\caption{Microservice architecture}
	\label{fig:microservices} 
\end{figure}

Microservices is a trending architectural style that aim to design complex systems as collections of fine grained and loosely coupled software artifacts called microservices; each microservice implements a small part or even a single function of the business logic of applications~\cite{Dragoni2017}. Their efficient loose coupling enables their development using different programming languages, use different database technologies, and be tested in isolation with respect to the rest of underlying systems. Microservices may communicate with each other directly using an HTTP resource API or indirectly by means of message brokers (see Figure~\ref{fig:microservices}). Microservices can either be deployed in virtual machines or lightweight containers. The use of containers for deploying microservices is preferred due to their simplicity, lower cost, and their fast initialization and execution.

Regarding software quality attributes, adopting microservices increases reusability and interoperability, enables scalability and enhances maintainability of complex software systems. Within adequate distributed platforms and technologies, microservices can easily be deployed, replicated, replaced, and destroyed independently without affecting systems availability. Moreover, implementing a single business capability per microservices allows their use in different applications and application domains. The main characteristics that differentiate microservices architectural style from monolithic and its ancestor service-oriented architectures is the smaller size, scalability and independence of each unit constituting a system.

Microservices are getting more attention and becoming adopted in industry. Currently, microservices are used by widely recognized companies such as Coca Cola, Amazon, eBay and NetFlix. Specifically, microservices are becoming more popular in software and IT service companies~\cite{Bogner2019}. 

Although the advantages brought by adopting microservice architectures in developing complex systems, security is one of the serious challenges that need to be tackled. Thus, there is an urgent need to identify and check current trends in overcoming security challenges in microservice architectures which is the aim of the present paper.%Those concerns are of much less interest in monolithic systems.

\subsection{Systematic mapping}
A systematic mapping is a kind of evidence-based software engineering (EBSE)~\cite{Kitchenham2015}. The aim of a systematic mapping is to provide an overview of a research area by building a classification scheme and structuring evidences on a research field. 

\begin{figure}[h]
	\centering
	\includegraphics[width=.6\textwidth]{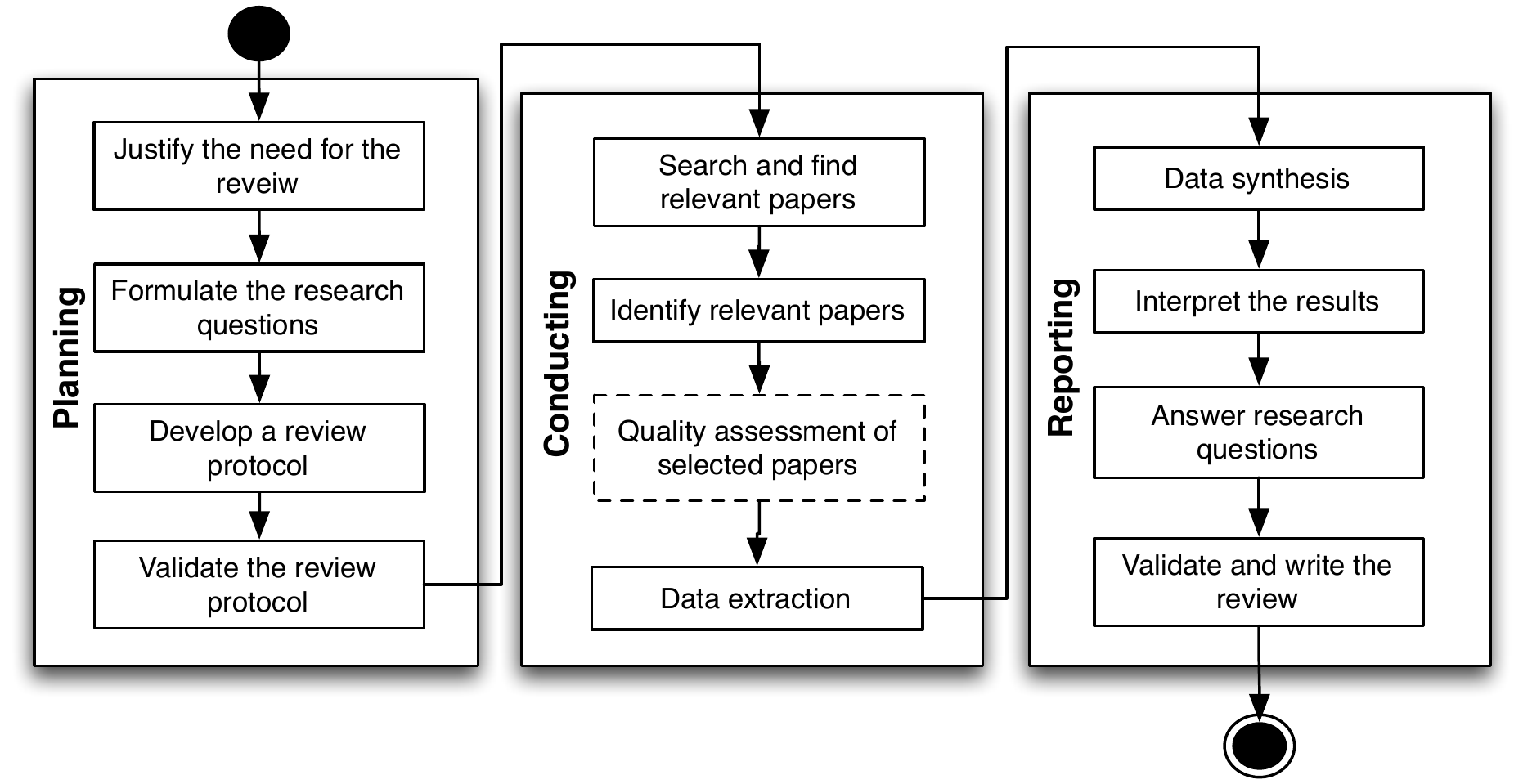}
	\caption{The overall process for elaborating systematic mapping}
	\label{fig:smsprocess} 
\end{figure}

Peterson et al.~\cite{Petersen2008,Petersen2015} has proposed an overall process for the elaboration of systematic mappings. The process is composed of three main steps: \textit{planning}, \textit{conducting} and \textit{reporting}. By planning, one can start by justifying the need and scope of the mapping, formulate the set of research questions, develop and validate a protocol specifying all the decisions relevant to conducting the mapping. The protocol includes the identification of search terms, search strategy, literature sources that need to be used to retrieve relevant papers, how and in which base found papers are selected and included in the mapping, what data need to be extracted from selected papers and how extracted data are synthesized and classified. By conducting, the initially validated protocol from the planning step is executed; thus, the identified sources are used to retrieve  papers, found papers are examined for relevance; useful data are then extracted from admitted papers and extracted data are synthesized and classified. By reporting, extracted data from primary papers are visualized, results are interpreted, research questions are answered and the mapping is validated and documented. Figure~\ref{fig:smsprocess} depicts the overall process for conducting systematic mapping studies as proposed in~\cite{Petersen2015}. The quality assessment step is depicted in Figure~\ref{fig:smsprocess} within a dashed line box since it is optional as stated in~\cite{Kitchenham2015,Petersen2015}.  

\section{Related Work}
\label{sec:rw}
We found in the literature several secondary studies (systematic reviews or mappings) dedicated to investigate the state-of-the-art of MSA in general. Surprisingly, few works are found focusing on security aspects in MSA. All found studies, except the work of Vale et al.~\cite{Vale2019}, are either platform or technology dependent investigations.   

Vale et al.~\cite{Vale2019} conducted a similar investigation to our study. They conducted a systematic mapping to reveal adopted security mechanisms for microservice-based systems. The study examined 26 papers published from November 2018 to March 2019. Vale et al.~\cite{Vale2019} focused only on security mechanisms and categorized the 18 identified mechanisms according to their focus, and classified validation techniques according to their nature. The study revealed that (1) authentication and authorization are the most frequently adopted mechanisms for securing microservices, (2) case studies and experiments are the most validation techniques used for security proposals, (3) absence of patterns for microservices-based systems security. Our study is broader and improves Vale et al.~\cite{Vale2019} work in several ways. In this study we include published papers since 2011. Moreover, besides security mechanisms, we also focus on identifying security threats and the applicability of proposed solutions regarding their execution platforms and architectural layers.  

Yu et al.~\cite{Yu2019} presented a survey on security in microservice-based fog applications. The survey included papers published between 2010 and 2017. The focus of Yu et al. ~\cite{Yu2019} was domain specific; they focused on determining security challenges and potential solutions of adopting microservices in fog computing. The security issues identified by the study concerns containers, data, and network vulnerabilities. They also proposed a solution for inter-service communication in fog applications. 

Monteiro et al.~\cite{Monteiro2018} identified a set of elements related to microservices implementation in cloud computing. They reported the same security aspects discussed by Yu et al.~\cite{Yu2019}. They concluded that availability and trustworthiness are the two major security requirements in MSA. 

Nkomo et al.~\cite{Nkomo2019} conducted a systematic review on practices that can be incorporated into the development process of microservice-based systems. The focus of Nkomo et al.~\cite{Nkomo2019} was to propose general guidelines where security can be tackled earlier when developing microservice-based applications putting much emphasis on microservices composition. They ended up with five security-focused development activities: (1) identify  security requirements of microservice composition, (2) adopt secure programming best practices, (3) validate security requirements and secure programming best practices, (4) secure configuration of runtime infrastructure, and (5) continuously monitor the behavior of microservices. Considering security as a primary concern throughout the life-cycle of microservice-based systems is mandatory; however, experiences show that all security threats cannot be identified earlier especially with the continuous evolving of technologies. 

Sultan et al.~\cite{Sultan2019} presented a survey on the security of containers; they identified main threats due to images, registries, orchestration, containers themselves, side channels and host OS risks. They distinguished two kinds of solutions to containers security: software-based and hardware-based solutions without further investigation of proposed solutions. Belair et al.~\cite{Belair2019} complements the work of Sultan et al.~\cite{Sultan2019} by proposing a taxonomy for containers' security proposals.  Belair et al.~\cite{Belair2019} focused on security solutions at the infrastructure level putting much emphasis on data transmission through virtualization. Three categories were identified: configuration-based, code-based and rule-based defense. They reported the fact that Linux security model (LSM), the powerful defense framework targeting Linux, cannot be easily adapted to containers to improve security. 

Compared with the works of Yu et al.~\cite{Yu2019} and Monteiro et al.~\cite{Monteiro2018}, our study is domain and platform independent and includes more recent endeavors with broader focus on proposed solutions to security threats. Compared with the works in~\cite{Sultan2019} and ~\cite{Belair2019}, we focus on our study on security issues concerning MSA in general and not only containers.

\section{Research methodology}
\label{sec:sms}
In this section we present the details of the protocol adopted for conducting this mapping study. Following the guidelines of Peterson et al.~\cite{Petersen2008}, a systematic mapping study should include the following primary steps: a definition of research questions, search for relevant papers, screening of found papers, propose or use an existing classification scheme, data extraction and studies mapping. In the sequel, we describe the details of each step.

\subsection{Research Questions}
\label{sec:questions}
The aim of this study is to identify the set of security vulnerabilities and how to be tackled in microservice-based systems. Thus, we formulate our research questions in light of the aims of our study and following the guidelines of Kuhrmann et al.~\cite{Kuhrmann2017}. This study is conducted with five main questions in mind: %each of which is broken down into a set of sub-questions:

\textbf{RQ1.} What are the most addressed security threats, risks, and vulnerabilities of microservices and microservice architectures and how they can be classified? This research question distinguishes the list of mostly treated vulnerabilities from those needing further investigations.

\textbf{RQ2.} What are existing approaches and techniques used for securing mircoservices and microservice architectures and how they can be classified? This research question provides an overview of existing approaches and techniques used for securing microservice-based systems.

\textbf{RQ3.} At what level of architecture the proposed techniques and approaches are applicable for securing microservcies? This research question indicates where security is applied highlighting the less focused levels of microservice architectures .  

\textbf{RQ4.} What domains or platforms are the focus of existing solutions for securing microservices and microservice architectures? This research question shows whether the focus of the proposed solutions is platform specific or platform independent. 

\textbf{RQ5.} What kind of evidence is given regarding the evaluation and validation of  proposed approaches and techniques for securing microservices and microservice architectures?  This research question evaluates the maturity of existing security techniques highlighting the set of empirical strategies used to validate proposed solutions.

\subsection{Search process}
Search string used in this study is designed to be generic and simple. It is constructed based on search terms concerned with \textit{population} and \textit{intervention} as suggested by Petticrew and Roberts in~\cite{Petticrew2006}. Population refers to the application area which is microservices and microservice architectures where intervention is security, vulnerabilities and attacks. Accordingly, final adopted search string is :

\begin{gather*} 
\text{("microservice" OR "micro-service" OR "micro service")}\\ 
\text{AND}\\ 
\text{("architecture" OR "design" OR "system" OR "structure")}\\
\text{AND}\\ 
\text{("security" OR "vulnerability" OR "attack")}
\end{gather*}

%\begin{block}
%\small{("microservice" OR "micro-service" OR "micro service") AND ("architecture" OR "design" OR "system" OR "structure") AND (“security" OR “vulnerability” OR “Attack”)}
%\end{block}

For retrieving relevant studies, we followed the guidelines of Kuhrmann et al.~\cite{Kuhrmann2017}. Thus, we adopted the use of the following online academic libraries:

\begin{itemize}
	\item IEEE Xplorer (\url{https://ieeexplore.ieee.org})
	\item ACM Digital Library (\url{https://dl.acm.org})
	\item SpringerLink (\url{https://link.springer.com})
	\item ScienceDirect (\url{https://www.sciencedirect.com/})
	\item Wiley Online Library (\url{https://onlinelibrary.wiley.com})
\end{itemize}

To avoid missing relevant studies, we complement our automatic search by conducting recursive backward and forward snowballing on selected studies as suggested by Wohllin~\cite{Wohlin2014,Wohlin2016}. By backward snowballing, we check the relevance of references in approved papers. By forward snowballing, we check the relevance of papers citing approved papers. The snowballing is recursively applied to each newly approved paper. Google Scholar is used as a sole source for forward snowballing. 

\subsection{Study selection process}
The set of retrieved papers by automatic search followed two screening stages. In the first stage, titles and abstracts were read to measure relevance. In the second stage, full texts of papers were examined to check if they meet our inclusion criteria. The list of all the papers are screened separately by the two authors; decisions are exchanged and conflicts are discussed and solved. Found papers from snowballing are also screened separately by the two authors before deciding whether to be included or excluded. 

\subsection{Inclusion and exclusion criteria}
The number of retrieved papers by online academic libraries is reduced by specifying a strict number of inclusion and exclusion criteria. In this study, only peer-reviewed papers from journals and conferences are included. The automatic search is conducted to cover all published papers since 2011 including early publications. The starting year 2011 is adopted since there was no consensus on the term microservice architectures prior 2011~\cite{Dragoni2017}. Only English written papers addressing security aspects or security solutions to microservices or microservice architectures are included. The full list of adopted inclusion and exclusion criteria are presented in Table~\ref{tab:inclusion} and Table~\ref{tab:exclusion} respectively.

\begin{table}[h]
	\centering
	\scriptsize{
		\caption{Inclusion criteria}
		\label{tab:inclusion}       % Give a unique label
		\begin{tabular}{lp{10.5cm}}
			\hline\noalign{\smallskip}
			ID & Criteria\\
			\noalign{\smallskip}\hline\noalign{\smallskip}
			I1 & papers published since 2011 including early publications\\
			I2 & papers written in English\\
			I3 & papers subject to peer reviews\\
			I4 & papers including studies conducted with security aspects of microservices or microservice architectures as their primary topics\\
			I5 & papers proposing frameworks, techniques, methods, or tools to secure microservices or microservice architectures\\
			I6 & papers presenting qualitative or quantitative evaluation of security techniques used for microservices or microservice architectures\\ 
			\noalign{\smallskip}\hline
	\end{tabular}}
\end{table}

\begin{table}[h]
	%table caption is above the table
	\centering
	\scriptsize{
		\caption{Exclusion criteria}
		\label{tab:exclusion}       % Give a unique label
		\begin{tabular}{lp{10.5cm}}
			\hline\noalign{\smallskip}
			ID & Criteria\\
			\noalign{\smallskip}\hline\noalign{\smallskip}
			E1 & papers addressing security in distributed platforms and technologies such as clouds without explicit referring to microservices\\ 
			E2 & papers describing general aspects of microservice architectures without putting much emphasis on the security issue\\
			E3 & tutorial papers and editorials\\
			E4 & papers presenting reviews, surveys or secondary studies on the security of microservices or microservice architectures\\ 
			E5 & books or book chapters, because they usually undergo little peer review and present general ideas already published in journals or conferences\\
			E6 & papers without full text available\\
			\noalign{\smallskip}\hline
	\end{tabular}}
\end{table}

\subsection{Data extraction process}
Following the guidelines of Peterson et al.~\cite{Petersen2008} a data extraction form is designed as illustrated in Table~\ref{tab:dataextraction}. Each paper is described in terms of its metadata such as year of publication, source and type. In addition, a set of required information for our analysis are extracted. These include the list of security threats or attacks addressed by the study,  proposed solutions, application level of proposed solutions, validation method and application platforms.

\begin{table}[h]
	\centering
	\scriptsize{
		\caption{Data extraction form}
		\label{tab:dataextraction}       % Give a unique label
		\begin{tabular}{lp{2.7cm}p{11cm}l}
			\hline\noalign{\smallskip}
			ID & Data item & Description&RQ\\
			\noalign{\smallskip}\hline\noalign{\smallskip}
			D1& Study ID & first author name + year\\ 
			D2& Year  & year of the publication\\
			D3& Source & source of the publication\\
			D4& Type & conference or journal paper\\ 
			D5& Category & analysis, solution proposal or case study \\
			D6& Threats & addressed security threats& RQ1\\
			D7& Source of Threats & internal or external & RQ1\\
			D8& Solution type & general protection measures, framework, technique, tool or methodology proposal& RQ2\\
			D9& Security mechanisms& set of security mechanisms proposed or used in the study&RQ2\\
			D10& Applicability level & architectural level where the security mechanism is applied& RQ3\\
			D11& Validation method & verification and validation techniques used to check the feasibility of the proposed solution & RQ5\\  
			D12& Domain/Platform & domains or platforms applicable for the proposed solution&RQ4\\
			\noalign{\smallskip}\hline
	\end{tabular}}
\end{table}

\subsection{Data synthesis}
We noticed a lack of a consensus on detailed taxonomies for security threats and security mechanisms; this prevents mapping all the selected studies to appropriate and distinct categories answering research questions RQ1 and RQ2. Moreover, due to the diversity of applications used in selected studies, their targeted platforms and used verification and validation techniques, it was necessary to properly categorize those studies answering RQ3, RQ4 and RQ5.   

For mapping properly all the selected studies to proper categories for each research question, we used our experiences and existing taxonomies~\cite{Monteiro2018,Yarygina2018a,OWSAP2017} in identifying categories and their relationships. We also used grounded theory~\cite{Strauss1998} as a complementary approach to generate missing categories from extracted data items. Specifically, we used open coding and selective coding to identify categories and their relationships with existing categories from D6, D9 and D10-11. In this study, grounded theory is used in an iterative process, where categories and subcategories are changed in each iteration until reaching a stability state.

\section{Results of the mapping}
\label{sec:results}
In this section we describe and detail the results of the mapping study answering the five research questions outlined in section~\ref{sec:questions}. 

\subsection{Overview of selected studies}
The search process is conducted in December 2019 and yielded 46 distinct papers published since 2011. The designed query is applied to the set of selected libraries. Table~\ref{tab:searchresults} shows the number of returned papers by each library.

\begin{table}[h]
	\centering
	\scriptsize{
		\caption{Number of studies returned by each repository.}
		\label{tab:searchresults}       % Give a unique label
		\begin{tabular}{lr}
			\hline\noalign{\smallskip}
			Repository & Search results\\
			\noalign{\smallskip}\hline\noalign{\smallskip}
			IEEE Xplorer & 50\\
			ACM Digital Library & 46\\
			SpringerLink & 678\\
			ScienceDirect & 255\\
			Wiley Online Library & 38\\
			\noalign{\smallskip}\hline
			Total & 1067\\
			\noalign{\smallskip}\hline
	\end{tabular}}
\end{table}

\begin{figure}[h]
	\centering
	\includegraphics[width=.3\textwidth]{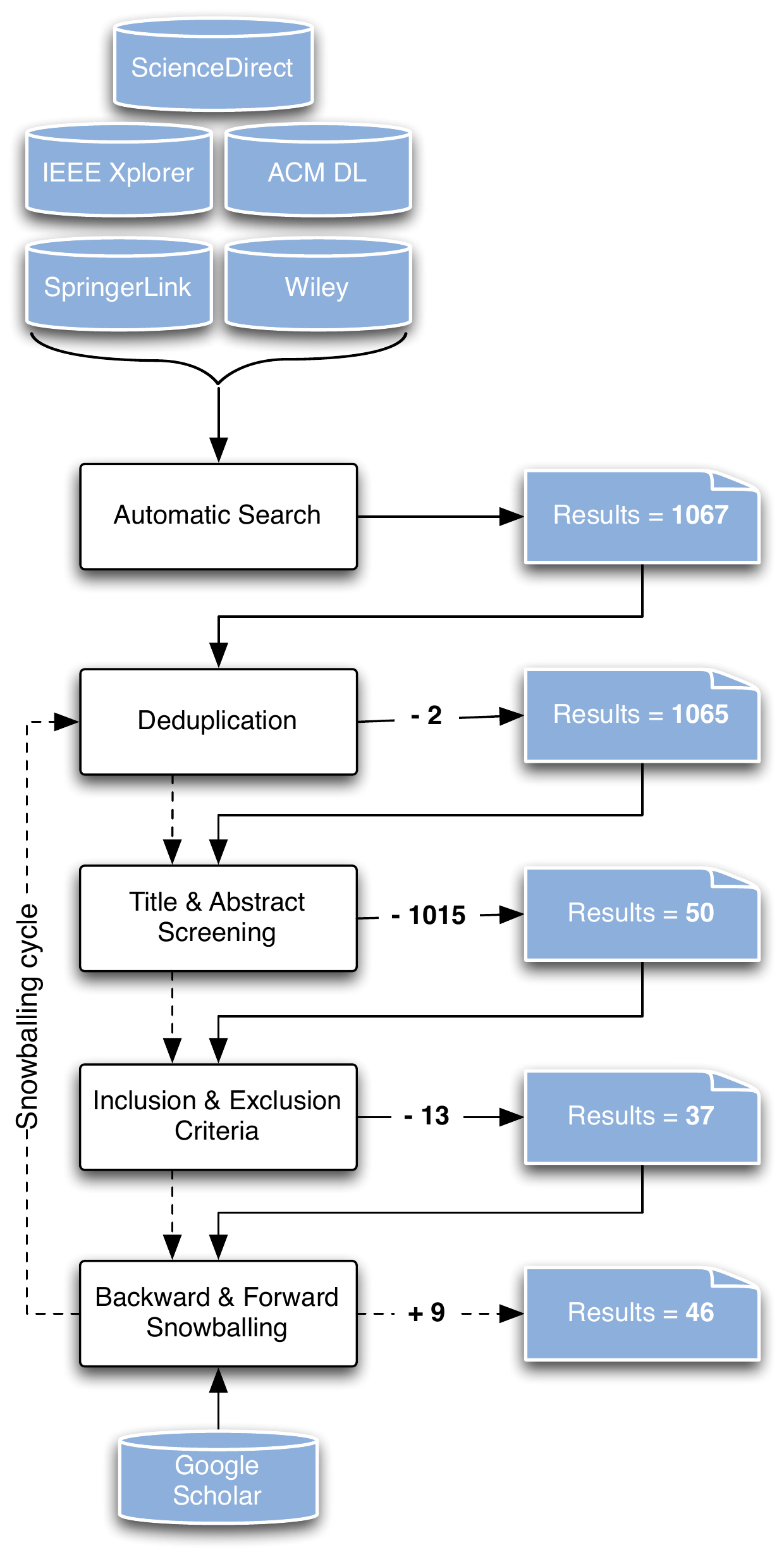}
	\caption{Paper selection process}
	\label{fig:search}       % Give a unique label
\end{figure}

The set of 1067 retrieved papers by the different search engines are gathered and duplicate papers are removed. This reduces the number to 1065. By screening titles and abstracts of remaining papers, 1015 papers are excluded for their irrelevance. After checking the inclusion and exclusion criteria, only 37 papers are approved. By conducting recursive backward and forward snowballing, 9 more papers are added. Two snowballing cycles are performed before reaching a steady state. In the first round, 7 new papers are included; in the second round, 2 more papers are added. Figure~\ref{fig:search} depicts the overall selection process. 

\begin{figure}[h]
	\centering
	\includegraphics[width=.53\textwidth]{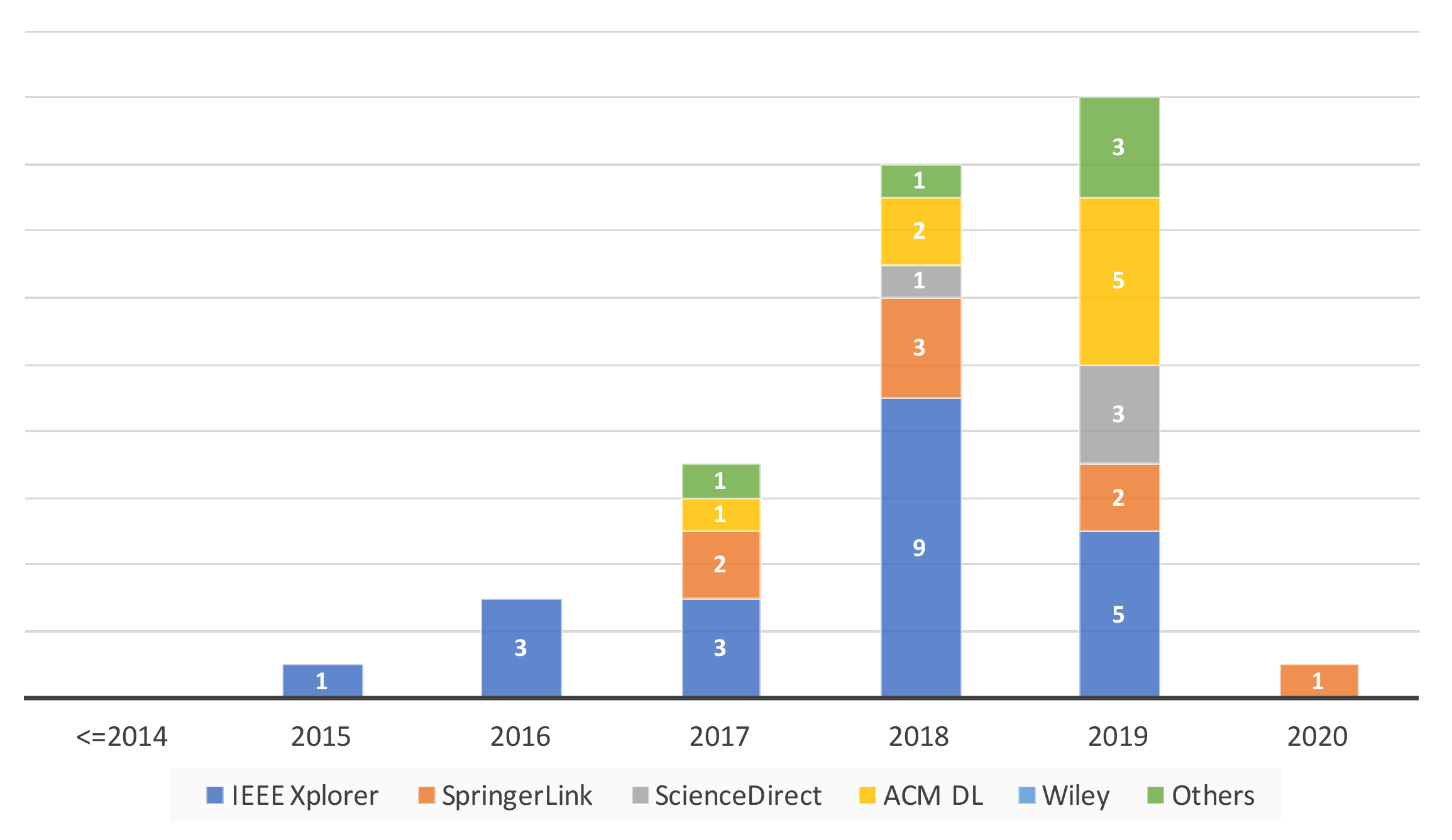}
	\caption{Distribution of selected studies by year and digital library}
	\label{fig:peryear}       % Give a unique label
\end{figure}

Figure~\ref{fig:peryear} shows the distribution of selected studies according to their publication year and source. We notice that, although the earlier emergence of MSA in 2011, the interest into securing microservices and microservice architectures is considered few years later and start getting more attentions since 2015. Figure~\ref{fig:peryear} also shows that the maximum number of publications come from IEEE Xplorer and none from Wiley meets our inclusion criteria.

Table~\ref{tab:studies} shows the complete list of selected studies with their year and type of publication and how they are found.

%\onecolumn{
%\lipsum[1-3]
\begin{table}[hpt]
	\centering
	\scriptsize{
		\caption{List of selected studies: C: Conference paper, J: Journal paper, A: Automatic search, S: Snowballing}
		\label{tab:studies}       % Give a unique label
		\begin{tabular}{lccccc}
			\hline\noalign{\smallskip}
			ID & Cite & Year & Type & Publisher & Search type\\
			\noalign{\smallskip}\hline\noalign{\smallskip}
			p1&\cite{Nkomo2019}&2019&C & Springer & A\\
			p2&\cite{Ahmadvand2018}&2018&C &Springer& A\\
			p3&\cite{Surantha2020}&2020&C &Springer& A\\
			p4&\cite{Brenner2017}&2017&C &Springer& A\\
			p5&\cite{Otterstad2017}&2017&C &Springer& A\\
			p6&\cite{Yarygina2018}&2018&C &Springer& A\\
			p7&\cite{Nehme2019a}&2019&C &Springer& A\\
			p8&\cite{Banati2018}&2018&C &IEEE& A\\
			p9&\cite{Nagothu2018}&2018&C &IEEE& A\\
			p10&\cite{Pahl2018}&2018&C&IEEE& A\\
			p11&\cite{Thanh2016}&2016&C&IEEE& A\\
			p12&\cite{Sun2015}&2015&C&IEEE& A\\
			p13&\cite{Buzachis2018}&2018&C&IEEE& A\\
			p14&\cite{George2017}&2017&C&IEEE& A\\
			p15&\cite{Ranjbar2017}&2017&C&IEEE& A\\
			p16&\cite{Ahmadvand2016}&2016&C&IEEE& A\\
			p17&\cite{Yarygina2018a}&2018&C&IEEE& A\\
			p18&\cite{Torkura2018}&2018&C&IEEE& A\\
			p19&\cite{Jin2019}&2019&J&IEEE& A\\
			p20&\cite{Gerking2019}&2019&C&IEEE& A\\
			p21&\cite{Osman2019}&2019&C&IEEE& A\\
			p22&\cite{Pahl2018a}&2018&C&IEEE& A\\
			p23&\cite{Ravichandiran2018}&2018&C&IEEE& A\\
			p24&\cite{Wen2019}&2019&J&IEEE& A\\
			p25&\cite{Lu2017}&2017&C&IEEE& S\\
			p26&\cite{Pahl2018b}&2018&C&IEEE& S\\
			p27&\cite{Nehme2019}&2019&J&IEEE& S\\
			p28&\cite{Fetzer2016}&2016&J&IEEE& S\\
			p29&\cite{Nguyen2019}&2019&J&JSW& S\\
			p30&\cite{He2017}&2017&J&IOP& S\\
			p31&\cite{Baker2019}&2019&C&CITRENZ& S\\
			p32&\cite{Salibindla2018}&2018&J&IJERT& S\\
			p33&\cite{Jander2019}&2019&J&IASKS& S\\
			p34&\cite{Torkura2018a}&2018&C&Springer& A\\
			p35&\cite{Chen2019}&2019&C&ACM& A\\
			p36&\cite{Torkura2017}&2017&C&ACM& A\\
			p37&\cite{Akkermans2018}&2018&C&ACM& A\\
			p38&\cite{Guija2018}&2018&C&ACM& A\\
			p39&\cite{Li2019}&2019&C&ACM& A\\
			p40&\cite{Marquez2019}&2019&C&ACM& A\\
			p41&\cite{Ibrahim2019}&2019&C&ACM& A\\
			p42&\cite{Stallenberg2019}&2019&C&ACM& A\\
			p43&\cite{Kramer2019}&2019&J&Science Direct& A\\
			p44&\cite{Jander2018}&2018&J&Science Direct& A\\
			p45&\cite{Abidi2019}&2019&J&Science Direct& A\\
			p46&\cite{Elsayed2019}&2019&J&Science Direct& A\\
			\noalign{\smallskip}\hline
	\end{tabular}}
\end{table}

\begin{figure}[hpt]
	\centering
	\includegraphics[width=.4\textwidth]{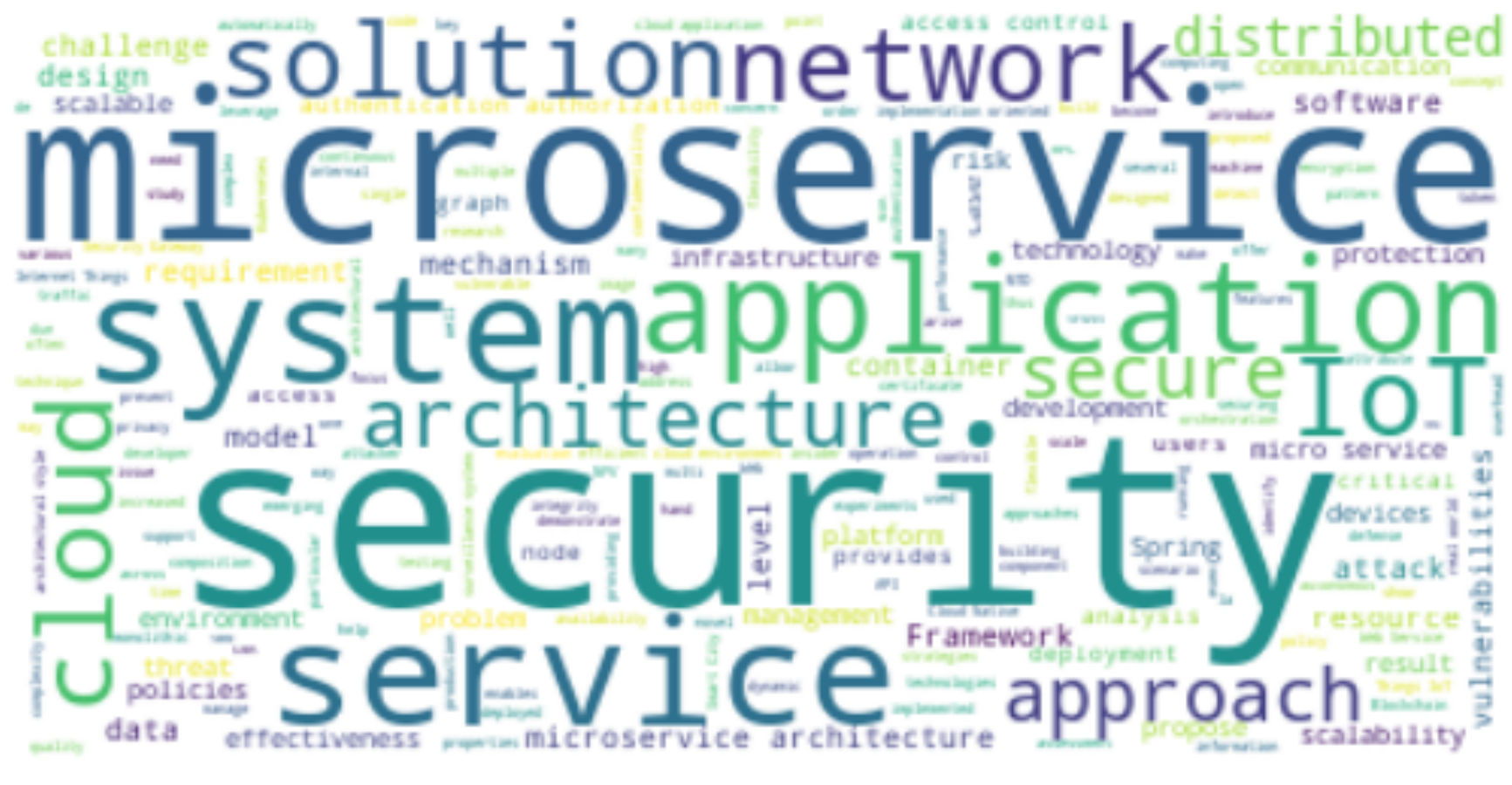}
	\caption{Keywords cloud of primary studies}
	\label{fig:wordcloud}       % Give a unique label
\end{figure}

Following the guidelines of Kuhrmann et al.~\cite{Kuhrmann2017}, we also experienced the use of Word Clouds to analyze the appropriateness of our result set of primary studies. Figure~\ref{fig:wordcloud} illustrates the most frequent words used in selected papers based on their titles and abstracts. The Figure shows that the most used words are microservice, security, application, architecture, system and service. Attacks, vulnerabilities and risks are rarely used in titles and abstracts.

\subsection{MSA security threats (RQ1)}
\begin{figure}[h]
	\centering
	\includegraphics[width=.4\textwidth]{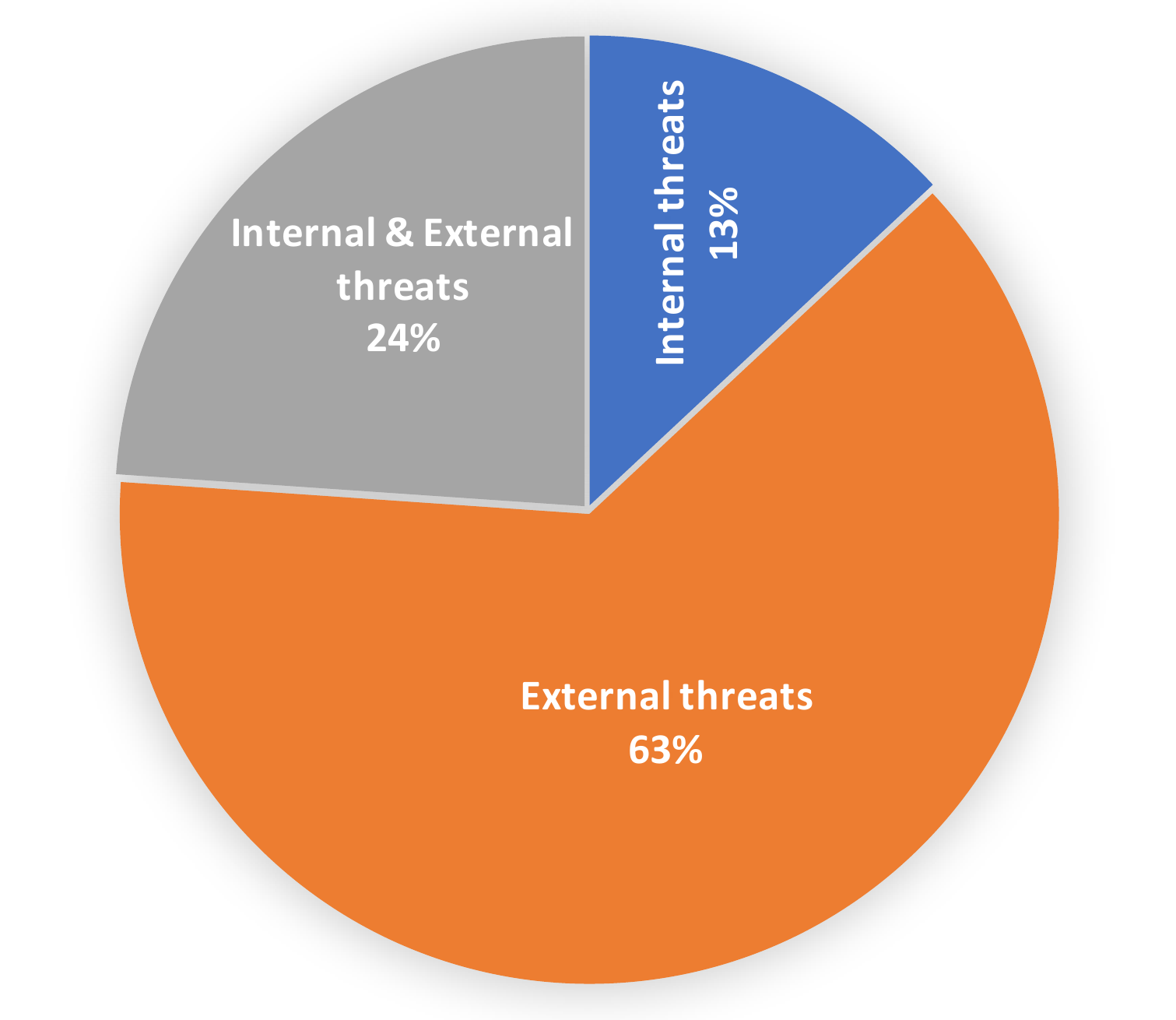}
	\caption{MSA security source of threats}
	\label{fig:threatsources}       % Give a unique label
\end{figure}

Microservice architecture as an emerging development paradigm in software engineering brings new security threats and vulnerabilities. These threats may come from insiders (i.e. internal attacks) or from outsiders (i.e. external attacks). For proper securing microservice-based systems, all threats, regardless of their origin, need to be detected and prevented using either available mitigation techniques or through proposing innovative solutions. In this study, we identify the focus of existing endeavors with respect to the source of threats (internal, external or both). Figure~\ref{fig:threatsources} depicts the distribution of identified and selected studies regarding the addressed source of threats. Figure~\ref{fig:threatsources} shows that 63\% of primary studies focus on external attacks, only 13\% focus on internal attacks and 24\% focus on both source of threats. This clearly indicates an unbalanced research focus towards external attacks.

Due to the plethora of taxonomies of security threats and lack of a consensus among their categorization, we adopted, a classification based on targets of attacks. Accordingly, threats in MSA can be classified into:

\begin{itemize}
	\item \textit{User-based attacks:} attacks where users are involved directly (i.e. malicious user actions) or indirectly (i.e. inadvertent insider actions).
	\item \textit{Data attacks:} threats targeting sensitive data that can be disclosed and manipulated by attackers.
	\item \textit{Infrastructure attacks:} attacks targeting MSA architectural elements and platforms such as monitors, discovery service, message broker, load balancer, etc.
	\item \textit{Software attacks:} threats involving code transformation or injection for malicious purposes.
\end{itemize}

Table~\ref{tab:secThreatCat} shows the set of MSA security threats addressed by primary studies grouped by category. The results revealed that unauthorized access, sensitive data exposure and compromising individual microservices are the most treated and addressed threats by contemporary studies. In addition, infrastructure attacks are the most diverse but with less addressed attacks in selected studies. 

\begin{table}[h]
	\scriptsize{
		\caption{MSA security threats per category}
		\label{tab:secThreatCat}       % Give a unique label
		\begin{tabular}{llr>{\raggedleft\arraybackslash}p{4.7cm}}
			\hline\noalign{\smallskip}
			Threats & \multicolumn{2}{c}{Percentile} & Studies\\
			\noalign{\smallskip}\hline\noalign{\smallskip}
			\rowcolor{lightgray}\textbf{User-based Attacks}  &\DrawPercentageBar{0.5870} & \textbf{58.70\%}&\\
			Brute Force Attack &\DrawPercentageBar{0.0435}& 4.35\%  &\cite{Nguyen2019,Baker2019}\\
			Cross-Site Request Forgery (CSRF) &\DrawPercentageBar{0.130}& 13.00\%   &\cite{Nehme2019a,Nguyen2019,Yarygina2018,Baker2019,Torkura2018a,Torkura2017}\\
			Spoofing& \DrawPercentageBar{0.0435} & 4.35\% &\cite{Yarygina2018,Pahl2018b}\\	
			Malicious Insider&\DrawPercentageBar{0.0435}& 4.35\%  &\cite{Wen2019, Nehme2019a}\\
			Unauthorized Access&\DrawPercentageBar{0.50}& 50.00\%  &\cite{Nkomo2019,Surantha2020,Nehme2019a,Banati2018,Nagothu2018,Thanh2016,Buzachis2018,George2017,Ranjbar2017,Yarygina2018,Pahl2018a,Lu2017,Pahl2018b,Nguyen2019,He2017,Baker2019,Salibindla2018,Akkermans2018,Guija2018,Li2019,Kramer2019,Jander2018,Abidi2019}\\
			Violate non-repudiation& \DrawPercentageBar{0.0217}& 2.17\%  &\cite{Ahmadvand2018}\\
			\rowcolor{lightgray}\textbf{Data Attacks} & \DrawPercentageBar{0.4783} & \textbf{47.83\%}&\\
			Eavesdropping&\DrawPercentageBar{0.0217} & 2.17\% &\cite{Kramer2019}\\
			Heartblead&\DrawPercentageBar{0.0435} & 4.35\% &\cite{Yarygina2018,Pahl2018b}\\
			Man in The Middle attack (MiTM)&\DrawPercentageBar{0.0652}& 6.52\%  &\cite{Ranjbar2017,Yarygina2018,Marquez2019}\\
			Padding Oracle On Downgraded Legacy Encryption attack (POODLE)&\DrawPercentageBar{0.0217}& 2.17\%  &\cite{Yarygina2018}\\
			Replay attack& \DrawPercentageBar{0.0652}& 6.52\%  &\cite{Ranjbar2017,Jander2019,Jander2018}\\
			Sensitive data exposure&\DrawPercentageBar{0.3043}& 30.43\%  &\cite{Ahmadvand2018,Brenner2017,Nagothu2018,Buzachis2018,George2017,Yarygina2018,Gerking2019,Wen2019,Lu2017,Fetzer2016,Li2019,Kramer2019,Abidi2019,Elsayed2019}\\
			Sniffing attack& \DrawPercentageBar{0.0652} & 6.52\% &\cite{Surantha2020,Nehme2019a,Yarygina2018}\\
			\rowcolor{lightgray}\textbf{Infrastructure Attacks}  &\DrawPercentageBar{0.4130} & \textbf{41.30\%}&\\
			Compromise containers &\DrawPercentageBar{0.1521}& 15.21\%  &\cite{Yarygina2018a,Ranjbar2017,Jin2019,Torkura2018a,Chen2019,Torkura2017, Ibrahim2019}\\
			Compromise virtual machines &\DrawPercentageBar{0.0870}& 8.70\%  &\cite{Yarygina2018a,Sun2015,Torkura2018a,Torkura2017}\\
			Compromise discovery service & \DrawPercentageBar{0.0435}& 4.35\%  &\cite{Nkomo2019,Yarygina2018}\\
			Compromise hypervisor &\DrawPercentageBar{0.0870}& 8.70\% &\cite{Brenner2017,Yarygina2018a,Yarygina2018,Fetzer2016}\\
			Compromise management interface&\DrawPercentageBar{0.0217}& 2.17\%  &\cite{Wen2019}\\
			Compromise network nodes& \DrawPercentageBar{0.0217}& 2.17\%  &\cite{Yarygina2018a}\\						
			Compromise operating systems &\DrawPercentageBar{0.0435}& 4.35\% &\cite{Brenner2017,Fetzer2016}\\			
			Downgrade attack& \DrawPercentageBar{0.0435}& 4.35\% &\cite{Ahmadvand2018,Akkermans2018}\\			
			Hardware backdoors& \DrawPercentageBar{0.0217}& 2.17\%  &\cite{Yarygina2018}\\			
			Malicious images& \DrawPercentageBar{0.0435} & 4.35\% &\cite{Yarygina2018,Ibrahim2019}\\			
			Malicious provider& \DrawPercentageBar{0.0217} & 2.17\% &\cite{Yarygina2018}\\					
			Misconfiguration& \DrawPercentageBar{0.0435} & 4.35\% &\cite{Yarygina2018,Pahl2018a}\\				
			Port scan attack& \DrawPercentageBar{0.0217}& 2.17\%  &\cite{Surantha2020}\\				
			Sandbox escape&\DrawPercentageBar{0.0217} & 2.17\%  &\cite{Yarygina2018}\\				
			Session hijacking& \DrawPercentageBar{0.0435}& 4.35\% &\cite{Nehme2019a,Yarygina2018}\\				
			Stress attack&\DrawPercentageBar{0.0435} & 4.35\%  &\cite{Ravichandiran2018,Wen2019}\\				
			Cold boot attack&\DrawPercentageBar{0.0217} & 2.17\%  &\cite{Brenner2017}\\	
			
			\rowcolor{lightgray}\textbf{Software Attacks}  &\DrawPercentageBar{0.500} & \textbf{50.00\%}&\\
			
			Code reuse attack &\DrawPercentageBar{0.0435}& 4.35\%  &\cite{Torkura2018,Marquez2019}\\
			Compromise microservices & \DrawPercentageBar{0.3261}& 32.61\%  &\cite{Otterstad2017,Yarygina2018a,Pahl2018,Sun2015,Yarygina2018,Jin2019,Osman2019,Pahl2018a,Pahl2018b,Jander2019,Torkura2018a,Torkura2017,Ibrahim2019,Jander2018,Elsayed2019}\\			
			Disrupt sensitive operation&\DrawPercentageBar{0.0217}& 2.17\%  &\cite{Ahmadvand2018}\\			
			Denial of Service (DoS)&\DrawPercentageBar{0.1739}& 17.39\%  &\cite{Ranjbar2017,Yarygina2018,Nguyen2019,Baker2019,Torkura2018a,Torkura2017,Stallenberg2019}\\	
			Injection&\DrawPercentageBar{0.1522}& 15.22\%   &\cite{Nehme2019a,Yarygina2018a,Nguyen2019,Baker2019,Torkura2018a,Torkura2017,Stallenberg2019}\\
			
			\noalign{\smallskip}\hline
	\end{tabular}}
\end{table}

Although IBM X-Force~\cite{IBM2016} reported that 60\% of all attacks were carried out by insiders, the study shows that only 13\% of primary studies focus on internal attacks. This is probably due to the fact that external threats are easier to be handled compared with internal threats. External threats are common in networking systems and can usually be identified and prevented by means of strong firewalls and intrusion detection systems; internal threats often requires considerable policy changes and continuous monitoring of internal traffics. This is owing to privileges awarded and sensitive data exposed to insiders. 

The diversity of attacks is due to the adoption of Zero Trust model~\cite{Kindervag2012} that suggests to afford no default trust to users, devices, applications, or packets; instead every action and entity need to be authenticated and authorized appropriately. Moreover, infrastructure attacks are less addressed due to their complexity since most attacks require low level solutions especially those related to hardware, nodes and operating systems. Attacks from other categories often require high level or software-based solutions that can easily be integrated into existing platforms and technologies. This justifies why software, user-based, and data attacks earned more attention than infrastructure attacks.  
Thus, we advocate for research studies that investigate threats caused by insiders in microservice-based applications. In addition, we suggest to investigate all OWASP identified vulnerabilities with their effects when adopting microservice architectures.

%\begin{figure}[h]
%	% Use the relevant command to insert your figure file.
%	% For example, with the graphicx package use
%	\centering
%	\includegraphics[width=.46\textwidth]{images/securitythreats3.pdf}
%	% figure caption is below the figure
%	\caption{MSA security threats per category}
%	\label{fig:st}       % Give a unique label
%\end{figure}

\subsection{Microservice security mechanisms (RQ2)}

Due to the diversity of proposed solutions, we classify MSA security mechanisms addressed in primary studies regarding the nature of their proposals as follows: 

\begin{itemize}
	\item \textit{General protection measures:} use of general security techniques to mitigate common known threats in MSA, or a set of general guidelines on choosing appropriate languages and technologies.
	\item \textit{Framework-based solutions:} architectural frameworks for MSA incorporating specific modules to handle some security aspects and mechanisms such as authorization, continuous monitoring, diagnosis.
	\item \textit{Technique-based solutions:} newly designed or adopted techniques from other domains to mitigate or prevent some security threats in MSA.
	\item \textit{Tool-based solutions:} newly developed tools implementing security measures.
	\item  \textit{Algorithm-based solutions:} new algorithms conceived for the detection or prevention of security threats.
	\item  \textit{Protocol-based solutions:} new protocols conceived for the protection of communications among the different MSA architectural elements.
	\item  \textit{Analysis:} experimentation, comparison or discussion of existing security mechanisms of MSA.
\end{itemize}

\begin{figure}[h]
	\centering
	\includegraphics[width=.6\textwidth]{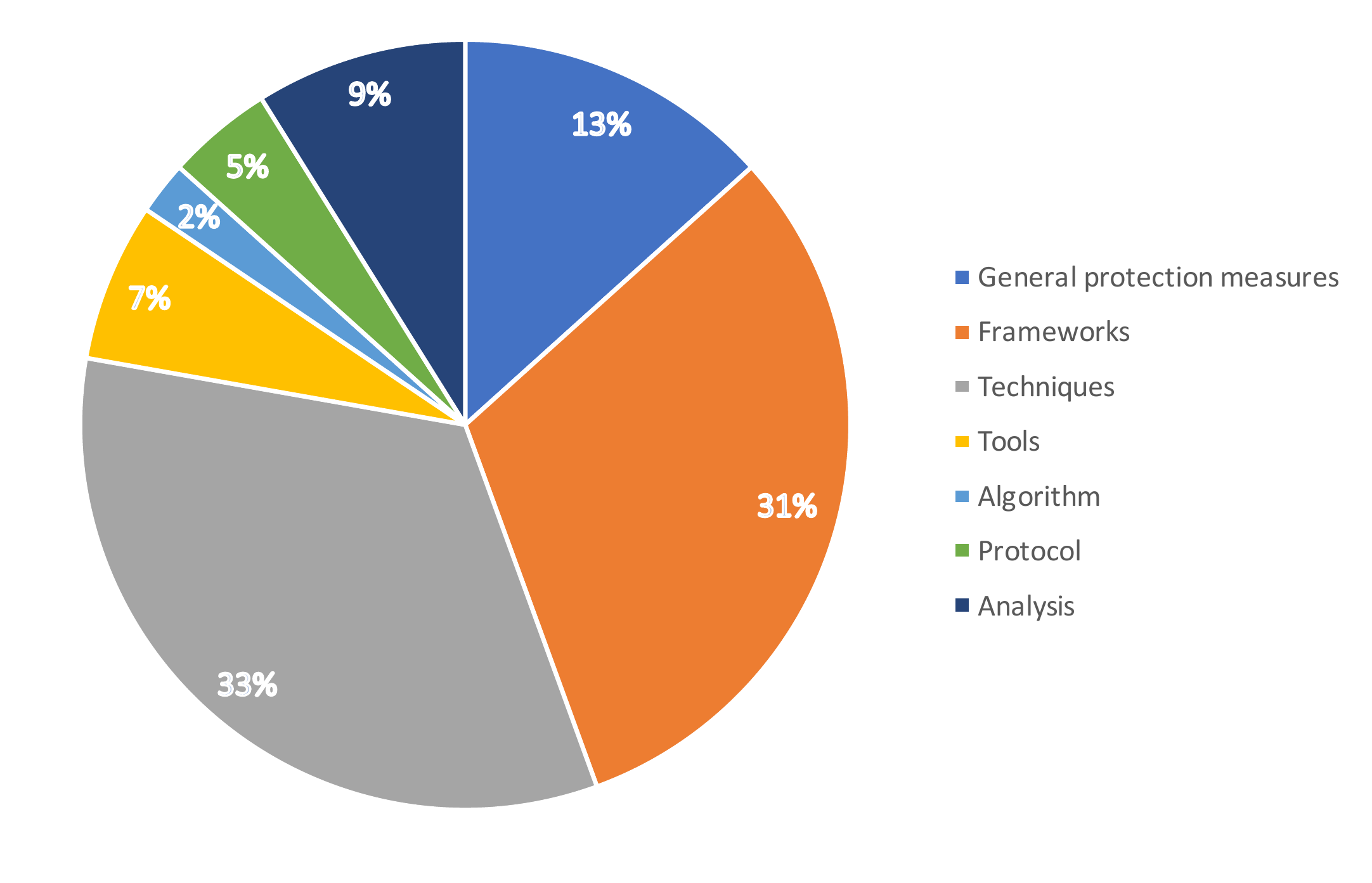}
	\caption{MSA security solutions}
	\label{fig:stypes}       % Give a unique label
\end{figure}

Our investigation (see Figure~\ref{fig:stypes}) shows that 33\% of the studies proposed new techniques for securing MSA and 31\% proposed framework-based solutions and 13\% proposed general protection measures. Few studies developed new tools, algorithms or protocols. Specifically, the authors of P17 have analyzed existing security mechanisms and proposed a framework based on the insights of the conducted analysis. 

Proposed solutions for securing microservices and microservice architectures can be classified into proposals for enforcing Access Control (i.e. authentication and/or authorization policies), auditing, mitigation and prevention:

\begin{itemize}
	\item \textit{Access Control - Authentication:} techniques used to verify the identity of users requiring access to MSA resources and data.
	\item \textit{Access Control - Authorization:} techniques used to check users' permissions for accessing specific MSA resources or data.
	\item \textit{Auditing:} techniques applied at runtime for discovering security gaps and may: (1) subsequently initiate appropriate measures or (2) simply report security breaches to relevant supervisory authority.
	\item \textit{Mitigation:} techniques that limit the damage of attacks when they appear. Mitigation techniques can be integrated into existing mircoservice-based systems.
	\item \textit{Prevention:} techniques that try to stop attacks from happening in the first place. Prevention techniques need to be considered when developing new mircoservice-based systems.
\end{itemize}

%\begin{figure}[hpt]
%	% Use the relevant command to insert your figure file.
%	% For example, with the graphicx package use
%	\centering
%	\includegraphics[width=.42\textwidth]{images/securitymechanisms.pdf}
%	% figure caption is below the figure
%	\caption{MSA security mechanism per category}
%	\label{fig:smech}       % Give a unique label
%\end{figure}

\begin{table}[h]
	\centering
	\scriptsize{
		\caption{MSA security mechanism per category}
		\label{tab:secMechCat}       % Give a unique label
		\begin{tabular}{llr>{\raggedleft\arraybackslash}p{4.5cm}}
			\hline\noalign{\smallskip}
			Security Mechanisms& \multicolumn{2}{c}{Percentile} & Studies\\
			\noalign{\smallskip}\hline\noalign{\smallskip}			
			\rowcolor{lightgray}\textbf{Authentication}  &\DrawPercentageBar{0.2826} & \textbf{28.26\%}&\\
			
			Centralized Access Control Manager&\DrawPercentageBar{0.0652}& 6.52\%  &\cite{Nehme2019a,Thanh2016,Ranjbar2017}\\			
			Certificates &\DrawPercentageBar{0.0652} & 6.52\% &\cite{Nkomo2019,Yarygina2018,Pahl2018b}\\			
			Open ID& \DrawPercentageBar{0.0652} & 6.52\% &\cite{Banati2018,Nehme2019,Guija2018}\\			
			Single Sign On (SSO)& \DrawPercentageBar{0.0435} & 4.35\% &\cite{Banati2018,He2017}\\			
			White-list HTTP/IP&\DrawPercentageBar{0.0435}& 4.35\%   &\cite{Nkomo2019,Salibindla2018}\\
			HIP exchange protocol &\DrawPercentageBar{0.0217} &2.17\%&\cite{Ranjbar2017}\\			
			J-PAKE protocol& \DrawPercentageBar{0.0435} & 4.35\% &\cite{Jander2019,Jander2018}\\
			Distribute sessions &\DrawPercentageBar{0.0217}& 2.17\% &\cite{He2017}\\
			HTTP signatures&\DrawPercentageBar{0.0217} & 2.17\% &\cite{Salibindla2018}\\

			\rowcolor{lightgray}\textbf{Authorization} & \DrawPercentageBar{0.2174} & \textbf{21.74\%}&\\
			
			Attribute Based Access Control (ABAC) &\DrawPercentageBar{0.0217} & 2.17\% &\cite{Thanh2016}\\
			Role Based Access Control (RBAC)&\DrawPercentageBar{0.0652}& 6.52\%  &\cite{Nehme2019a,Jander2019,Guija2018}\\
			R/W Permission to message broker&\DrawPercentageBar{0.0217}& 2.17\%   &\cite{Nkomo2019}\\		
			OAuth 2&\DrawPercentageBar{0.1739}& 17.39\%  &\cite{Nehme2019a,Banati2018,Thanh2016,Nehme2019,Nguyen2019,Baker2019,Salibindla2018,Guija2018}\\
			
			\rowcolor{lightgray}\textbf{Authentication \& Authorization} & \DrawPercentageBar{0.2391}& \textbf{23.91\%}&\\	
			JSON Web Token (JWT) &\DrawPercentageBar{0.1739}& 17.39\%  &\cite{Nkomo2019,Banati2018,Yarygina2018,George2017,Nehme2019,He2017,Salibindla2018,Guija2018}\\	
			Firewalls & \DrawPercentageBar{0.0870}& 8.70\% &\cite{Surantha2020,Pahl2018,Thanh2016,Nehme2019}\\
			
			\rowcolor{lightgray}\textbf{Auditing}  &\DrawPercentageBar{0.4348}& \textbf{43.48\%}&\\
			
			Continuous monitoring&\DrawPercentageBar{0.3478}& 34.78\%  &\cite{Ahmadvand2018,Surantha2020,Otterstad2017,Yarygina2018a,Pahl2018,Sun2015,Osman2019,Pahl2018a,Ravichandiran2018,Nehme2019,Torkura2018a,Chen2019,Torkura2017,Li2019,Ibrahim2019,Abidi2019}\\
			Scan container images &\DrawPercentageBar{0.0217}& 2.17\%  &\cite{Nkomo2019}\\			
			Static/Dynamic code analysis&\DrawPercentageBar{0.0870}& 8.70\%  &\cite{Yarygina2018,Fetzer2016,Li2019,Elsayed2019}\\
			Machine learning &\DrawPercentageBar{0.0652}& 6.52\%   &\cite{Pahl2018,Pahl2018a,Chen2019}\\			
			Intrusion detection&\DrawPercentageBar{0.0435}& 4.35\%  &\cite{Yarygina2018a,Nehme2019}\\
			
			\rowcolor{lightgray}\textbf{Mitigation}  &\DrawPercentageBar{0.1087}& \textbf{10.87\%}&\\
			
			Roll-back/Restart microservices &\DrawPercentageBar{0.0217} & 2.17\% &\cite{Yarygina2018a}\\
			Scale up/down N-variant microservices&\DrawPercentageBar{0.0217}& 2.17\%  &\cite{Yarygina2018a}\\
			Short-lived tokens& \DrawPercentageBar{0.0217}& 2.17\%  &\cite{Nehme2019a}\\
			Diversification&\DrawPercentageBar{0.0435}& 4.35\%  &\cite{Yarygina2018a,Torkura2018}\\
			IP shuffling &\DrawPercentageBar{0.0217} & 2.17\% &\cite{Jin2019}\\
			Live migration &\DrawPercentageBar{0.0435} & 4.35\% &\cite{Jin2019,Osman2019}\\
			Deception &\DrawPercentageBar{0.0217} & 2.17\% &\cite{Osman2019}\\
			Isolation of suspicious microservices  &\DrawPercentageBar{0.0217} & 2.17\% &\cite{Yarygina2018a}\\
			
			\rowcolor{lightgray}\textbf{Prevention}  &\DrawPercentageBar{0.3478} & \textbf{34.78\%}&\\
			Blockchain technology & \DrawPercentageBar{0.0435} & 4.35\%  &\cite{Nagothu2018,Buzachis2018}\\			
			Encryption&\DrawPercentageBar{0.1087}& 10.87\%  &\cite{Ahmadvand2018,Ranjbar2017,Yarygina2018a,Lu2017,Kramer2019}\\
			Hardware Security Module (HMS) &\DrawPercentageBar{0.0217} & 2.17\% &\cite{Yarygina2018}\\			
			Least privilege & \DrawPercentageBar{0.0217} & 2.17\%&\cite{Yarygina2018}\\
			No shared memory access&\DrawPercentageBar{0.0217} & 2.17\% &\cite{Yarygina2018}\\			
			Proper design& \DrawPercentageBar{0.0652}& 6.52\% &\cite{Ahmadvand2016,Gerking2019,Marquez2019}\\			
			Secure languages&  \DrawPercentageBar{0.0435}& 4.35\% &\cite{Yarygina2018,Akkermans2018}\\			
			Smart contracts&\DrawPercentageBar{0.0652}& 6.52\%  &\cite{Nagothu2018,Thanh2016,Buzachis2018}\\
			TLS protocol&\DrawPercentageBar{0.0652} & 6.52\% &\cite{Brenner2017,George2017,Yarygina2018}\\			
			SGX technology with enclaves&\DrawPercentageBar{0.0652}& 6.52\%  &\cite{Brenner2017,Yarygina2018,Fetzer2016}\\
			
			\noalign{\smallskip}\hline
	\end{tabular}}
\end{table}

Table~\ref{tab:secMechCat} shows the list of proposed solutions mapped into our  classification with the proportion rate for each proposal with respect to the total set of primary studies. The results show that much emphasis is put on proposing auditing techniques (43.48\%), enforcing authentication and/or authorization (39.13\%\footnote{This value is calculated by collecting all the papers from authorization and/or authentication categories and removing duplicates; the obtained number is divided by the total number of papers.}), and prevention (34.78\%), where less attention is being paid to mitigation (10.87\%).

The much emphasis put on authentication and authorization techniques is defensible. In fact, authentication and authorization are basic security mechanisms to any secure system. They form a front defense line in the protection of the different microservice architecture elements (i.e. individual microservices, API gateway, containers, microservice registry, etc.). However, studies considering authentication and authorization are less innovative since they propose combination of existing techniques and standards. For example, the authors of P8 proposed a combination of OAuth 2.0, JWT, Open ID and SSO used by a special authentication and authorization orchestrator. 

Besides general continuous monitoring and code analysis, audition proposals are showing the integration of artificial intelligence techniques such as machine learning and self-learning algorithms (P10, P22, P35). Those techniques are based on runtime analysis of user and/or microservice behaviors and (semi-)automatically take predefined actions in reaction to suspicious behaviors.

Most, if not all, proposals for mitigation are Moving Target Defense-based solutions (MTD)~\cite{Zhuang2014}. The idea behind MTD is to continuously perform transformation of system components and configurations preventing attackers from acquiring knowledge about target systems to be used to initiate harmful attacks. This includes, periodically update or restart microservices, IP shuffling, and live migration of microservices. Specifically, the authors of P21 proposed deception through live cloning and sandboxing of suspicious containers respecting the same network overloading and performance to deceive attackers . 

Prevention proposals are the most diverse techniques. They varies from using physical computing devices such as Hardware Security Module (HMS), powerful techniques and technologies such as encryption and Blockchain into adopting software design decisions such as using secure programming languages and smart contracts.

Due to the lower rate of mitigation techniques and their applicability to existing microservice-based systems, we advocate more research studies on mitigation techniques.

\subsubsection{Micorservice security application levels (RQ3)}
The adoption of MSA as an architectural design of distributed applications introduces security vulnerabilities in different architectural layers. Thus, security measures need to be taken in every layer of MSA. In this study, we distinguish the following layers:

\begin{itemize}
	\item \textit{Microservice:} Individual microservices are the mainstays of MSA, those micorservices can be blocked or compromized through injection of malicious code. Thus, security measures to adopt only trusted micorservices and protect them from internal and external attacks need to be taken. 
	\item \textit{Composition:} Connections among microservices can be broken. Moreover, compromising a single microservice may affect the security of the whole system due to the insecure configuration options for individual microservices, their locations and inter-connections. Several security measures need to be taken at this level to secure the overall architecture of microservice-based systems. 
	\item \textit{API:} Finely tuned attacks on APIs can bypass traditional security measures provided by API gateways. Hence, assets can be accessed and controlled by malicious users. Appropriate security measures should be taken at API gateways to avoid such vulnerabilities.
	\item \textit{Communication:} Data exchanged between microservices through event-buses can be intercepted and altered by malicious insiders. Thus, securing communication channels between microservices is mandatory for securing microservice-based systems.
	\item \textit{Deployment:} Containers holding micorservices can also be sources of vulnerabilities. Containers can be compromised through gaining an unauthorized access or deriving vulnerabilities from using images from untrusted sources. Thus, appropriate security measures should also be taken at this level.
	\item \textit{Soft-Infrastructure:} Infrastructure vulnerabilities are lower level vulnerabilities that can affect practically every software entity running on the network including monitors, registries, message brokers, load balancers and other orchestrators. Thus, introducing techniques at this level to guarantee the security of the diverse software network entities and the safety of their configuration is of higher importance.
	\item \textit{Hard-Infrastructure:} Hardware components are also vulnerable to attacks. Attackers may use bugs and backdoors intentionally or unintentionally introduced at manufacturing~\cite{Mavroudis2017} to initiate attacks. These vulnerabilities need to be tackled by introducing appropriate error and backdoor detection mechanisms. 
\end{itemize}

\begin{table}[h]
	\centering
	\scriptsize{
		\caption{MSA security application levels}
		\label{tab:secAppLev}       % Give a unique label
		\begin{tabular}{llr>{\raggedleft\arraybackslash}p{6.7cm}}
			\hline\noalign{\smallskip}
			Application Layer & \multicolumn{2}{c}{Percentile} & Studies\\
			\noalign{\smallskip}\hline\noalign{\smallskip}
			
			Microservice &\DrawPercentageBar{0.2}& 20.00\%  &\cite{Otterstad2017,Yarygina2018a,Torkura2018,Lu2017,Pahl2018b,Nehme2019,Li2019,Stallenberg2019,Elsayed2019}\\
			Composition & \DrawPercentageBar{0.0667}& 6.67\%  &\cite{Ahmadvand2016,Gerking2019,Nehme2019}\\
			API &\DrawPercentageBar{0.3556}& 35.56\% &\cite{Nkomo2019,Ahmadvand2018,Nehme2019a,Banati2018,Nagothu2018,Thanh2016,Buzachis2018,George2017,Ranjbar2017,Yarygina2018,Lu2017,Nguyen2019,He2017,Baker2019,Salibindla2018,Guija2018}\\
			Communication &\DrawPercentageBar{0.2222}& 22.22\%  &\cite{Brenner2017,Banati2018,Nagothu2018,Buzachis2018,George2017,Ranjbar2017,Yarygina2018,Jander2019,Guija2018,Jander2018}\\
			Deployment &\DrawPercentageBar{0.1333}& 13.33\%   &\cite{Brenner2017,Jin2019,Wen2019,Lu2017,Fetzer2016,Abidi2019}\\
			Soft-Infrastructure&\DrawPercentageBar{0.6889}& 68.89\%  &\cite{Nkomo2019,Ahmadvand2018,Surantha2020,Otterstad2017,Yarygina2018a,Nehme2019a,Banati2018,Nagothu2018,Pahl2018,Thanh2016,Sun2015,George2017,Ranjbar2017,Yarygina2018,Torkura2018,Osman2019,Pahl2018a,Ravichandiran2018,Pahl2018b,Nehme2019,Fetzer2016,He2017,Torkura2018a,Chen2019,Torkura2017,Akkermans2018,Guija2018,Li2019,Ibrahim2019,Kramer2019,Elsayed2019}\\
			Hard-Infrastructure&\DrawPercentageBar{0.0889} & 8.89\% &\cite{Brenner2017,Jin2019,Osman2019,Fetzer2016}\\
			\noalign{\smallskip}\hline
	\end{tabular}}
\end{table}

Table~\ref{tab:secAppLev} shows the distribution of solutions provided by primary studies per their application layers. The study revealed that much emphasis are put to conceive solutions applicable at soft-infrastructure and API gateways where less attention is being paid to composition and hard-infrastructure layers. 

Although, microservices are the mainstay of MSA, securing individual microservices is not getting a higher rate. The less interest in hard-infrastructure solutions is defensible due to their complexity and cost-intensive compared with soft-infrastructure based solutions. However, individual microservices, their composition and communication should have much attention than that has been revealed. Specifically, communication protection is of a high importance regarding the huge number and nature of transmitted data in the communication channels.

\subsubsection{MSA security mechanisms target platforms (RQ4)}
The identified papers in this study are classified regarding the target platforms and applications for their proposed solutions. Table~\ref{tab:secPlatform} shows that 34.78\% of papers proposed MSA security solutions that work for different platforms; closer proportion is found for  solutions to deal with securing microservices in the cloud platform. Few studies proposed platform specific solutions such as 5G platform, IoT, Web applications, kubernetes platforms and Springer framework.

\begin{table}[h]
	\centering
	\scriptsize{
		\caption{MSA security application platforms}
		\label{tab:secPlatform}       % Give a unique label
		\begin{tabular}{llr>{\raggedleft\arraybackslash}p{6cm}}
			\hline\noalign{\smallskip}
			Applications/ Platforms & \multicolumn{2}{c}{Percentile} & Studies\\
			\noalign{\smallskip}\hline\noalign{\smallskip}
			
			IoT applications &\DrawPercentageBar{0.1304}& 13.04\%  &\cite{Pahl2018,George2017,Pahl2018a,Lu2017,Pahl2018b,Akkermans2018}\\
			
			Cloud platforms & \DrawPercentageBar{0.2826}& 28.26\%  &\cite{Brenner2017,Banati2018,Thanh2016,Sun2015,Ranjbar2017,Jin2019,Ravichandiran2018,Wen2019,Fetzer2016,Torkura2018a,Torkura2017,Kramer2019,Elsayed2019}\\
			Osmotic computing &\DrawPercentageBar{0.0217} & 2.17\%&\cite{Buzachis2018}\\
			
			Container-based platforms &\DrawPercentageBar{0.1087}& 10.87\%  &\cite{Torkura2018,Jin2019,Osman2019,Chen2019,Ibrahim2019}\\
			
			Web applications  & \DrawPercentageBar{0.0652} & 6.52\% &\cite{Ravichandiran2018,Stallenberg2019,Abidi2019}\\
			
			Springer platform &\DrawPercentageBar{0.0435} & 4.35\% &\cite{Nguyen2019,Baker2019}\\
			
			Kubernetes platform &\DrawPercentageBar{0.0217} & 2.17\%&\cite{Surantha2020}\\
			
			5G platform&\DrawPercentageBar{0.0217} & 2.17\% &\cite{Guija2018}\\
			
			Independent&\DrawPercentageBar{0.3478}& 34.78\%  &\cite{Nkomo2019,Ahmadvand2018,Otterstad2017,Yarygina2018a,Nehme2019a,Nagothu2018,Ahmadvand2016,Yarygina2018,Gerking2019,Nehme2019,He2017,Salibindla2018,Jander2019,Li2019,Marquez2019,Jander2018}\\
			\noalign{\smallskip}\hline
	\end{tabular}}
\end{table}

Cloud-focused and platform independent solutions are found within a higher rates, 34.78\% and 28.26\%, respectively. The interest to cloud computing is understandable due to different facilities provided to companies by adopting MSA for developing their applications. Adopting MSA for developing applications in the cloud allow companies to integrate existing legacy systems, to grow with demands and to use up-to-date and intuitive interfaces. Solutions provided for IoT applications are also getting more attentions due to the specificity and the growing needs to those applications in the market. 

\subsubsection{Micorservice security V\&V methods (RQ5)}
For validating the proposed solutions, we distinguished the use of several verification and validation approaches: %use case based testing, performance analysis, case study based verification, formal verification, complexity measures, qualitative comparison and measuring adhoc metrics. 
\begin{itemize}
	\item \textit{Validation by simulation:} this includes: (1) use of simulated lab environments or testbeds for testing proposed designs, (2) simulation of attacks with check bypassing of proposed security measures.
	\item \textit{Manual testing:} this includes: (1) use case-based testing, (2) emulate attacks, and (3) reconfigure-testing cycles.
	\item \textit{Performance analysis:} this is performed by measuring overheads, latency, throughput, memory storage, CPU usage, response time, and traffic measurement.
	\item \textit{Qualitative analysis:} compare or verify and validate a set of qualitative requirements. These include: causing single point to failure, complexity of cracking, complexity of implementation and potential inherent bottleneck.
	\item \textit{Quantitative analysis:} comparing the proposed solution with similar proposals using quantitative metrics such as session sustainability, and popular machine learning metrics.
	\item \textit{Adhoc metrics:} proposing specific metrics for the evaluation of proposals. For example, P18 proposed a diversification index as a security measure to validate the proposal. Authors of P21 used a quality of deception metric to evaluate the proposed deception mechanism.
	\item \textit{Case study based validation:} use case studies to validate the feasibility of the proposed solution.
	\item \textit{Proof of concept (POC):} develop a prototype to demonstrate the feasibility of the proposed solution.
	\item \textit{Tool-based testing:} use unit-based testing tools such as IntelliJ IDEA\footnote{IntelliJ IDEA: \url{https://www.jetbrains.com/fr-fr/idea/}}.
	\item \textit{Formal verification:} use model checkers or theorem provers to check the validity of specified properties.
	\item \textit{Complexity measuring:} estimate temporal complexity of proposed algorithms implementing solutions. 
	\item \textit{Methodology based analysis:} use predefined and well-known methodologies such as OWASP risk rating, attack surface or security risk comparison. 	
\end{itemize}

%\begin{figure}[h]
%	% Use the relevant command to insert your figure file.
%	% For example, with the graphicx package use
%	\centering
%	\includegraphics[width=.48\textwidth]{images/validation.pdf}
%	% figure caption is below the figure
%	\caption{MSA security verification and validation methods}
%	\label{fig:validation}       % Give a unique label
%\end{figure}

\begin{table}[h]
	\centering
	\scriptsize{
		\caption{MSA security verification and validation methods}
		\label{tab:validation}       % Give a unique label
		\begin{tabular}{llr>{\raggedleft\arraybackslash}p{5.5cm}}
			\hline\noalign{\smallskip}
			V\&V methods & \multicolumn{2}{c}{Percentile} & Studies\\
			\noalign{\smallskip}\hline\noalign{\smallskip}
			
			Simulation & \DrawPercentageBar{0.0652} & 6.52\% &\cite{Surantha2020,George2017,Pahl2018b}\\
			
			Manual Testing  & \DrawPercentageBar{0.0870} & 8.70\% &\cite{Osman2019,Ravichandiran2018,Baker2019,Abidi2019}\\
			
			Performance analysis & \DrawPercentageBar{0.3913} & 39.13\% &\cite{Brenner2017,Yarygina2018a,Nehme2019a,Pahl2018,Sun2015,George2017,Ranjbar2017,Yarygina2018,Jin2019,Osman2019,Pahl2018a,Ravichandiran2018,Pahl2018b,Fetzer2016,Torkura2017,Akkermans2018,Kramer2019,Elsayed2019}\\
			
			Qualitative analysis & \DrawPercentageBar{0.1087} & 10.87\%&\cite{Otterstad2017,Banati2018,He2017,Salibindla2018,Kramer2019}\\
			
			Quantitative analysis &  \DrawPercentageBar{0.0870} & 8.70\%&\cite{Ranjbar2017,Elsayed2019,Pahl2018a,Chen2019}\\
			
			Adhoc metrics  &\DrawPercentageBar{0.0652} & 6.52\%&\cite{Torkura2018,Osman2019,Pahl2018a}\\
			
			Case study based validation  &\DrawPercentageBar{0.2609} & 26.09\% &\cite{Ahmadvand2018,Nagothu2018,Thanh2016,George2017,Ahmadvand2016,Lu2017,Jander2019,Li2019,Marquez2019,Ibrahim2019,Stallenberg2019,Jander2018}\\
			
			Proof of concept (POC)  &\DrawPercentageBar{0.0217} & 2.17\%&\cite{Nguyen2019}\\
			
			Tool-based testing  &\DrawPercentageBar{0.1087} & 10.87\%&\cite{Baker2019,Torkura2018a,Guija2018,Marquez2019,Elsayed2019}\\
			
			Formal verification &\DrawPercentageBar{0.0217} & 2.17\% &\cite{Gerking2019}\\
			
			Complexity measuring & \DrawPercentageBar{0.0652} & 6.52\%&\cite{Yarygina2018a,Osman2019,Wen2019}\\
			
			Methodology-based validation & \DrawPercentageBar{0.0435} & 4.35\%&\cite{Torkura2018,Wen2019}\\
			
			\noalign{\smallskip}\hline
	\end{tabular}}
\end{table}

Table~\ref{tab:validation} shows that performance analysis and case study based validation are the most adopted techniques for verification and validation. Formal verification, POC and methodology-based analysis are the least used methods for validation. This is due to the nature of proposed solutions. Formal verification can only be adopted when formal specification of systems and their properties are described (only P20), where complexity measures are adopted for algorithmic based solutions (only P6). Validation by simulation and adhoc metrics are found used in equal measure with a rate equal to 6.52\%. Most used simulation methods used simulated lab environments or testbeds. Only P3 used simulated attacks to evaluate the proposed technique. Note that most examined studies used more than one validation techniques and 2 studies (P17 and P27) proposed solutions without using any mentioned validation technique. Instead, they discussed the details of proposed solutions and claimed that they are sufficient enough to mitigate the addressed security threat(s). 

\section{An ontology for securing MSA}
\label{sec:ontology}
\begin{figure}[h]
	\centering
	\includegraphics[width=1\textwidth]{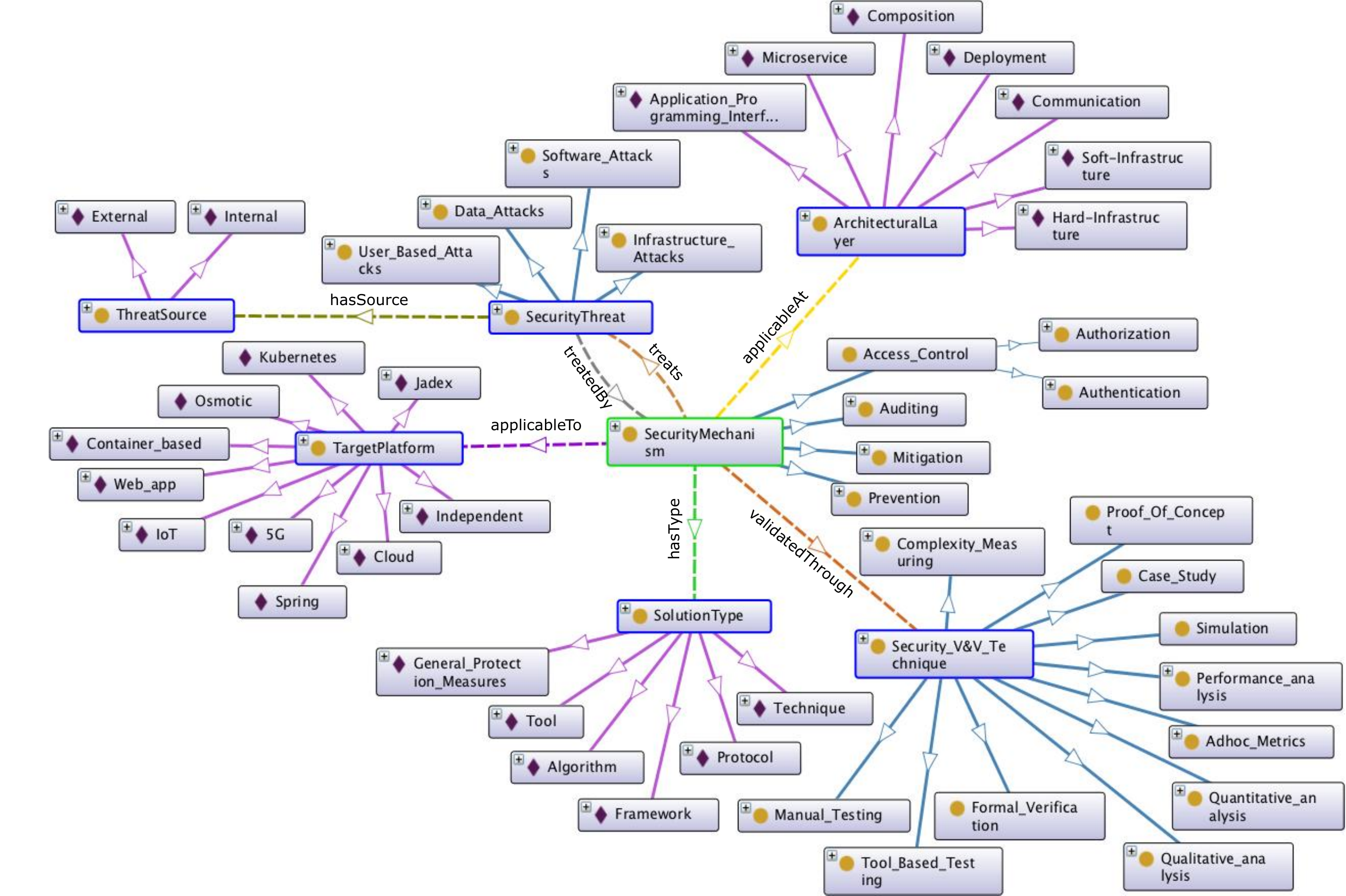}
	\caption{Ontology for MSA security}
	\label{fig:ontology}       % Give a unique label
\end{figure}

Making the results of our study practical and extendable and due to the static nature of taxonomies, we propose an ontology-based representation of  our results. Within an ontology, one can describe relationships among ontology concepts and individuals. In our study, relationships between security threats and security mechanisms are mandatory. In addition, mapping security mechanisms to their applicability architectural levels and platforms is necessary. Figure~\ref{fig:ontology} describes the overall MSA ontology retrieved from this study. The ontology is developed using Protégé~\footnote{https://protege.stanford.edu/} and its consistency and coherence are checked using the Protégé Debugger option. The main concepts of the ontology are:

\begin{enumerate}
	\item \textit{SecurityMechnism}: describes the set of security mechanisms proposed for MSA that have been retrieved by this study.
	\item \textit{SecurityThreat}: describes the set of security threats threatening microservices and microservice architecture that have been retrieved by this study.
	\item \textit{ArchitecturalLayer}: describes the different architectural layers of MSA.
	\item \textit{TargetPlatform}: describes the set of platforms addressed by the security mechanisms retrieved by this study. For platform independent solutions, an individual named \textit{Independent} is added to this concept class. 
	\item \textit{SolutionType}: describes the solution type for each proposed mechanism.
	\item \textit{ThreatSource}: describes the source of each threat: \textit{internal} or \textit{external}.
	\item \textit{Security\_V\&V\_Technique}: describes the verification and validation techniques used for recognized security mechanisms.
\end{enumerate}

Relationships between classes  are reflected through Protégé object properties  presented in Table~\ref{tab:properties}.  

\begin{table}[h]
	\centering
	\scriptsize{
		\caption{Relationships between ontology classes}
		\label{tab:properties}       % Give a unique label
		\begin{tabular}{lll}
			\hline\noalign{\smallskip}
			Object Properties & Domain & Rang\\
			\noalign{\smallskip}\hline\noalign{\smallskip}
			applicableAt & SecurityMechanism&ArchitecturalLayer\\
			applicableTo & SecurityMechanism&TargetPlatform\\
			hasSource & SecurityThreat&ThreatSource\\
			hasType & SecurityMechanism&SolutionType\\
			treatedBy & SecurityThreat&SecurityMechanism\\
			treats & SecurityMechanism&SecurityThreat\\
			validatedThrough & SecurityMechanism&Security\_V\&V\_Technique\\
			\noalign{\smallskip}\hline
	\end{tabular}}
\end{table}

For the usability of the ontology, OWL-DL queries can be used to investigate the ontology structure. Table~\ref{tab:queries} describes some of useful queries described in OWL-DL. In Table~\ref{tab:queries}, Q1 can be used to retrieve the list of security mechanisms applied to deal with \textit{Unauthorized\_access} threat. Q2 returns the list of security threats that can be alleviated by continuous monitoring. Q3 returns the list of security mechanisms applied at individual microservices. Finally, Q4 returns the list of internal threats recognized in this study. The overall ontology is available at \url{https://github.com/hannousse/MSASecurity} in an OWL format.

\begin{table}[h]
	\caption{OWL-DL Query samples}
	\label{tab:queries}       % Give a unique label
	\centering
	\scriptsize{
		\begin{tabular}{ll}
			\hline\noalign{\smallskip}
			
			ID & Query \\
			\noalign{\smallskip}\hline\noalign{\smallskip}
			Q1 & \text{SecurityMechanism \textbf{and} (treats \textbf{value} Unauthorized\_Access)}\\
			Q2 & \text{SecurityThreat \textbf{and} (treatedBy \textbf{value} Continuous\_Monitoring)}\\
			Q3 & \text{SecurityMechanism \textbf{and} (applicableAt \textbf{value} Micorservice)}\\
			Q4 & \text{SecurityThreat \textbf{and} (hasSource \textbf{value} Internal)}\\
			\noalign{\smallskip}\hline
	\end{tabular}}
\end{table}

\section{Threats to validity}
\label{sec:validity}
In this section we discuss the threats to the validity and how we mitigated their effects on the obtained results.

An internal validity threat to our study concerns the identification of primary studies from the large set of papers found in the literature. For these sake, we adopted the guidelines of Kuhrmann et al.~\cite{Kuhrmann2017} for the selection of search engines. To void bias to search engines, we completed our search by snowballing technique~\cite{Wohlin2016} over already identified papers. The use of several iterations of the snowballing technique allowed the identification of nine more relevant papers in which five were not indexed by the selected search engines. For ensuring the inclusion of high quality papers, we adopted a set of strict inclusion and exclusion criteria that accept only peer-reviewed journal and conference papers for their completeness and sufficient results. However, since only papers explicitly referring to microservices or microservice architectures were included, some papers focusing on securing the deployment layer, specifically Docker containers~\cite{Merkel2014} were omitted. 

A conclusion validity threat to our study concerns the adoption of taxonomies for security threats and mechanisms. In fact, several taxonomies are investigated~\cite{Monteiro2018,Yarygina2018a, OWSAP2017}, however, none of those taxonomies enable the proper classification of all the identified studies. Thus, we used open and selective coding from grounded theory~\cite{Strauss1998} and we adopted a classification based on deeper analysis of the focus and the proposed solutions of identified papers. Some of the categories of our classifications are already used in existing taxonomies, some they are either used as they are or adapted to fulfill the context of our study.  

\section{Conclusion}
\label{sec:conclusion}
In this study, we conducted a systematic mapping on securing microservices focusing on threats,  nature, applicability platforms, and validation techniques of security proposals. The study examined 46 papers published since 2011. The results revealed that unauthorized access, sensitive data exposure and compromising individual microservices are the most treated and addressed threats by contemporary studies. The results also revealed that auditing, enforcing access control, and prevention based solutions are the most proposed security mechanisms. Additionally, we found that most proposed solutions are applicable at soft-infrastructure layer of MSA. Our study shows that 34.78\% of papers proposed MSA security solutions that work for different platforms, the same proportion is noticed for cloud-based solutions. Finally, we found that most verification and validation methods were based on performance analysis, and case studies. We also proposed and made available of an ontology summarizing and gathering the retrieved results. the proposed ontology can be used as a guide to developers about already recognized threats and security mechanisms for MSA.  
We noticed that most addressed threats are well-known for other architectural styles and few are concerned directly with MSA. Specifically, compromising individual microservices that can radically lead to a chain defection in MSA. Moreover, continuous monitoring became very popular among MSA designers to prevent possibly future threats. Encryption remain the most used technique facing sensitive data exposure. Regarding noticed unbalanced research focus on external attacks and prevention techniques, we advocate more studies studying internal attacks and proposing mitigation techniques. Moreover, more studies are suggested for treating 
individual microservice and communication layers vulnerabilities.


\begin{thebibliography}{10}
	
	\bibitem{Yarygina2018a}
	T.~{Yarygina} and A.~H. {Bagge}.
	\newblock Overcoming security challenges in microservice architectures.
	\newblock In {\em 2018 IEEE Symposium on Service-Oriented System Engineering
		(SOSE)}, pages 11--20, March 2018.
	
	\bibitem{Baskarada2018}
	Sa{\v s}a Ba{\v s}karada, Vivian Nguyen, and Andy Koronios.
	\newblock Architecting microservices: Practical opportunities and challenges.
	\newblock {\em Journal of Computer Information Systems}, pages 1--9, 2018.
	
	\bibitem{Dragoni2017}
	Nicola Dragoni, Saverio Giallorenzo, Alberto~Lluch Lafuente, Manuel Mazzara,
	Fabrizio Montesi, Ruslan Mustafin, and Larisa Safina.
	\newblock {\em Microservices: Yesterday, Today, and Tomorrow}, chapter~12,
	pages 195--216.
	\newblock Springer International Publishing, Cham, 2017.
	
	\bibitem{Alshuqayran2016}
	Nuha Alshuqayran, Nour Ali, and Roger Evans.
	\newblock A systematic mapping study in microservice architecture.
	\newblock In {\em 2016 IEEE 9th International Conference on Service-Oriented
		Computing and Applications (SOCA)}, pages 44--51. IEEE, 2016.
	
	\bibitem{Bogner2019}
	J.~{Bogner}, J.~{Fritzsch}, S.~{Wagner}, and A.~{Zimmermann}.
	\newblock Microservices in industry: Insights into technologies,
	characteristics, and software quality.
	\newblock In {\em 2019 IEEE International Conference on Software Architecture
		Companion (ICSA-C)}, pages 187--195, 2019.
	
	\bibitem{Kitchenham2015}
	Barbara~Ann Kitchenham, David Budgen, and Pearl Brereton.
	\newblock {\em Evidence-Based Software Engineering and Systematic Reviews}.
	\newblock Chapman \& Hall/CRC, 2015.
	
	\bibitem{Petersen2008}
	Kai Petersen, Robert Feldt, Shahid Mujtaba, and Michael Mattsson.
	\newblock Systematic mapping studies in software engineering.
	\newblock In {\em Proceedings of the 12th International Conference on
		Evaluation and Assessment in Software Engineering}, EASE'08, pages 68--77,
	Swindon, UK, 2008.
	
	\bibitem{Petersen2015}
	Kai Petersen, Sairam Vakkalanka, and Ludwik Kuzniarz.
	\newblock Guidelines for conducting systematic mapping studies in software
	engineering: An update.
	\newblock {\em Information and Software Technology}, 64:1--18, 2015.
	
	\bibitem{Vale2019}
	Anelis~Pereira Vale, Gast{\'o}n M{\'a}rquez, Hern{\'a}n Astudillo, and
	Eduardo~B Fernandez.
	\newblock Security mechanisms used in microservices-based systems: A systematic
	mapping.
	\newblock In {\em XLV Latin American Computing Conference}, pages 1--10, 2019.
	
	\bibitem{Yu2019}
	Dongjin Yu, Yike Jin, Yuqun Zhang, and Xi~Zheng.
	\newblock A survey on security issues in services communication of
	microservices-enabled fog applications.
	\newblock {\em Concurrency and Computation: Practice and Experience},
	31(22):e4436, 2019.
	\newblock e4436 cpe.4436.
	
	\bibitem{Monteiro2018}
	Luciano de~Aguiar~Monteiro, Washington Henrique~Carvalho Almeida,
	Raphael~Rodrigues Hazin, Anderson~Cavalcanti de~Lima, Sahra Karolina~Gomes
	e~Silva, and Felipe~Silva Ferraz.
	\newblock A survey on microservice security--trends in architecture, privacy
	and standardization on cloud computing environments.
	\newblock {\em International Journal on Advances in Security},
	11(3-4):201--213, 2018.
	
	\bibitem{Nkomo2019}
	Peter Nkomo and Marijke Coetzee.
	\newblock Software development activities for secure microservices.
	\newblock In Sanjay Misra, Osvaldo Gervasi, Beniamino Murgante, Elena Stankova,
	Vladimir Korkhov, Carmelo Torre, Ana Maria~A.C. Rocha, David Taniar,
	Bernady~O. Apduhan, and Eufemia Tarantino, editors, {\em Poceedings of
		International Conference on Computational Science and Its Applications, ICCSA
		2019}, pages 573--585, Cham, 2019. Springer International Publishing.
	
	\bibitem{Sultan2019}
	S.~{Sultan}, I.~{Ahmad}, and T.~{Dimitriou}.
	\newblock Container security: Issues, challenges, and the road ahead.
	\newblock {\em IEEE Access}, 7:52976--52996, 2019.
	
	\bibitem{Belair2019}
	Maxime B{\'e}lair, Sylvie Laniepce, and Jean-Marc Menaud.
	\newblock Leveraging kernel security mechanisms to improve container security:
	A survey.
	\newblock In {\em Proceedings of the 14th International Conference on
		Availability, Reliability and Security}, ARES '19, pages 76:1--76:6, New
	York, NY, USA, 2019. ACM.
	
	\bibitem{Kuhrmann2017}
	Marco Kuhrmann, Daniel~M{\'e}ndez Fern{\'a}ndez, and Maya Daneva.
	\newblock On the pragmatic design of literature studies in software
	engineering: an experience-based guideline.
	\newblock {\em Empirical Software Engineering}, 22(6):2852--2891, 2017.
	
	\bibitem{Petticrew2006}
	Mark Petticrew and Helen Roberts.
	\newblock {\em Systematic Reviews in the Social Sciences: A Practical Guide}.
	\newblock John Wiley \& Sons, Ltd, 2006.
	
	\bibitem{Wohlin2014}
	Claes Wohlin.
	\newblock Guidelines for snowballing in systematic literature studies and a
	replication in software engineering.
	\newblock In {\em Proceedings of the 18th International Conference on
		Evaluation and Assessment in Software Engineering}, EASE '14, pages 1--10,
	New York, NY, USA, 2014. Association for Computing Machinery.
	
	\bibitem{Wohlin2016}
	Claes Wohlin.
	\newblock Second-generation systematic literature studies using snowballing.
	\newblock In {\em Proceedings of the 20th International Conference on
		Evaluation and Assessment in Software Engineering}, EASE '16, pages 1--6, New
	York, NY, USA, 2016. Association for Computing Machinery.
	
	\bibitem{OWSAP2017}
	OWASP.
	\newblock Owasp top 10: The ten most critical web application security risks.
	\newblock Technical report, OWASP Foundation, 2017.
	
	\bibitem{Strauss1998}
	Anselm~L. Strauss and Juliet~M. Corbin.
	\newblock {\em Basics of qualitative research: techniques and procedures for
		developing grounded theory}.
	\newblock Sage Publications, Thousand Oaks, Calif, 1998.
	
	\bibitem{Ahmadvand2018}
	Mohsen Ahmadvand, Alexander Pretschner, Keith Ball, and Daniel Eyring.
	\newblock Integrity protection against insiders in microservice-based
	infrastructures: From threats to a security framework.
	\newblock In Manuel Mazzara, Iulian Ober, and Gwen Sala{\"u}n, editors, {\em
		Proceedings of Federation of International Conferences on Software
		Technologies: Applications and Foundations}, pages 573--588, Cham, 2018.
	Springer International Publishing.
	
	\bibitem{Surantha2020}
	Nico Surantha and Felix Ivan.
	\newblock Secure kubernetes networking design based on zero trust model: A case
	study of financial service enterprise in indonesia.
	\newblock In Leonard Barolli, Fatos Xhafa, and Omar~K. Hussain, editors, {\em
		Proceedings of International Conference on Innovative Mobile and Internet
		Services in Ubiquitous Computing}, pages 348--361, Cham, 2020. Springer
	International Publishing.
	
	\bibitem{Brenner2017}
	Stefan Brenner, Tobias Hundt, Giovanni Mazzeo, and R{\"u}diger Kapitza.
	\newblock Secure cloud micro services using intel sgx.
	\newblock In Lydia~Y. Chen and Hans~P. Reiser, editors, {\em Proceedings of
		IFIP International Conference on Distributed Applications and Interoperable
		Systems}, pages 177--191, Cham, 2017. Springer International Publishing.
	
	\bibitem{Otterstad2017}
	Christian Otterstad and Tetiana Yarygina.
	\newblock Low-level exploitation mitigation by diverse microservices.
	\newblock In Flavio De~Paoli, Stefan Schulte, and Einar Broch~Johnsen, editors,
	{\em Proceedings of the 6th IFIP WG 2.14 European Conference on
		Service-Oriented and Cloud Computing}, pages 49--56, Cham, 2017. Springer
	International Publishing.
	
	\bibitem{Yarygina2018}
	Tetiana Yarygina and Christian Otterstad.
	\newblock A game of microservices: Automated intrusion response.
	\newblock In Silvia Bonomi and Etienne Rivi{\`e}re, editors, {\em Proceedings
		of IFIP International Conference on Distributed Applications and
		Interoperable Systems}, pages 169--177, Cham, 2018. Springer International
	Publishing.
	
	\bibitem{Nehme2019a}
	Antonio Nehme, Vitor Jesus, Khaled Mahbub, and Ali Abdallah.
	\newblock Fine-grained access control for microservices.
	\newblock In Nur Zincir-Heywood, Guillaume Bonfante, Mourad Debbabi, and
	Joaquin Garcia-Alfaro, editors, {\em Proceedings of the International
		Symposium on Foundations and Practice of Security}, pages 285--300, Cham,
	2019. Springer International Publishing.
	
	\bibitem{Banati2018}
	A.~{B{\'a}n{\'a}ti}, E.~{Kail}, K.~{Kar{\'o}czkai}, and M.~{Kozlovszky}.
	\newblock Authentication and authorization orchestrator for microservice-based
	software architectures.
	\newblock In {\em 2018 41st International Convention on Information and
		Communication Technology, Electronics and Microelectronics (MIPRO)}, pages
	1180--1184, May 2018.
	
	\bibitem{Nagothu2018}
	D.~{Nagothu}, R.~{Xu}, S.~Y. {Nikouei}, and Y.~{Chen}.
	\newblock A microservice-enabled architecture for smart surveillance using
	blockchain technology.
	\newblock In {\em 2018 IEEE International Smart Cities Conference (ISC2)},
	pages 1--4, Sep. 2018.
	
	\bibitem{Pahl2018}
	M.~{Pahl}, F.~{Aubet}, and S.~{Liebald}.
	\newblock Graph-based iot microservice security.
	\newblock In {\em NOMS 2018 - 2018 IEEE/IFIP Network Operations and Management
		Symposium}, pages 1--3, April 2018.
	
	\bibitem{Thanh2016}
	{Tran Quang Thanh}, S.~{Covaci}, T.~{Magedanz}, P.~{Gouvas}, and
	A.~{Zafeiropoulos}.
	\newblock Embedding security and privacy into the development and operation of
	cloud applications and services.
	\newblock In {\em 2016 17th International Telecommunications Network Strategy
		and Planning Symposium (Networks)}, pages 31--36, Sep. 2016.
	
	\bibitem{Sun2015}
	Y.~{Sun}, S.~{Nanda}, and T.~{Jaeger}.
	\newblock Security-as-a-service for microservices-based cloud applications.
	\newblock In {\em 2015 IEEE 7th International Conference on Cloud Computing
		Technology and Science (CloudCom)}, pages 50--57, Nov 2015.
	
	\bibitem{Buzachis2018}
	A.~{Buzachis} and M.~{Villari}.
	\newblock Basic principles of osmotic computing: Secure and dependable
	microelements (mels) orchestration leveraging blockchain facilities.
	\newblock In {\em 2018 IEEE/ACM International Conference on Utility and Cloud
		Computing Companion (UCC Companion)}, pages 47--52, Dec 2018.
	
	\bibitem{George2017}
	V.~M. {George} and Q.~H. {Mahmoud}.
	\newblock Claimsware: A claims-based middleware for securing iot services.
	\newblock In {\em 2017 IEEE 41st Annual Computer Software and Applications
		Conference (COMPSAC)}, volume~1, pages 649--654, July 2017.
	
	\bibitem{Ranjbar2017}
	A.~{Ranjbar}, M.~{Komu}, P.~{Salmela}, and T.~{Aura}.
	\newblock Synaptic: Secure and persistent connectivity for containers.
	\newblock In {\em 2017 17th IEEE/ACM International Symposium on Cluster, Cloud
		and Grid Computing (CCGRID)}, pages 262--267, May 2017.
	
	\bibitem{Ahmadvand2016}
	M.~{Ahmadvand} and A.~{Ibrahim}.
	\newblock Requirements reconciliation for scalable and secure microservice
	(de)composition.
	\newblock In {\em 2016 IEEE 24th International Requirements Engineering
		Conference Workshops (REW)}, pages 68--73, Sep. 2016.
	
	\bibitem{Torkura2018}
	K.~A. {Torkura}, M.~I.~H. {Sukmana}, and A.~V. D.~M. {Kayem}.
	\newblock A cyber risk based moving target defense mechanism for microservice
	architectures.
	\newblock In {\em 2018 IEEE Intl Conf on Parallel Distributed Processing with
		Applications, Ubiquitous Computing Communications, Big Data Cloud Computing,
		Social Computing Networking, Sustainable Computing Communications
		(ISPA/IUCC/BDCloud/SocialCom/SustainCom)}, pages 932--939, Dec 2018.
	
	\bibitem{Jin2019}
	H.~{Jin}, Z.~{Li}, D.~{Zou}, and B.~{Yuan}.
	\newblock Dseom: A framework for dynamic security evaluation and optimization
	of mtd in container-based cloud.
	\newblock {\em IEEE Transactions on Dependable and Secure Computing}, pages
	1--12, 2019.
	
	\bibitem{Gerking2019}
	C.~{Gerking} and D.~{Schubert}.
	\newblock Component-based refinement and verification of information-flow
	security policies for cyber-physical microservice architectures.
	\newblock In {\em 2019 IEEE International Conference on Software Architecture
		(ICSA)}, pages 61--70, March 2019.
	
	\bibitem{Osman2019}
	A.~{Osman}, P.~{Bruckner}, H.~{Salah}, F.~H.~P. {Fitzek}, T.~{Strufe}, and
	M.~{Fischer}.
	\newblock Sandnet: Towards high quality of deception in container-based
	microservice architectures.
	\newblock In {\em ICC 2019 - 2019 IEEE International Conference on
		Communications (ICC)}, pages 1--7, May 2019.
	
	\bibitem{Pahl2018a}
	M.~{Pahl} and F.~{Aubet}.
	\newblock All eyes on you: Distributed multi-dimensional iot microservice
	anomaly detection.
	\newblock In {\em 2018 14th International Conference on Network and Service
		Management (CNSM)}, pages 72--80, Nov 2018.
	
	\bibitem{Ravichandiran2018}
	R.~{Ravichandiran}, H.~{Bannazadeh}, and A.~{Leon-Garcia}.
	\newblock Anomaly detection using resource behaviour analysis for autoscaling
	systems.
	\newblock In {\em 2018 4th IEEE Conference on Network Softwarization and
		Workshops (NetSoft)}, pages 192--196, June 2018.
	
	\bibitem{Wen2019}
	Z.~{Wen}, T.~{Lin}, R.~{Yang}, S.~{Ji}, r.~{Ranjan}, A.~{Romanovsky}, C.~{Lin},
	and J.~{Xu}.
	\newblock Ga-par: Dependable microservice orchestration framework for
	geo-distributed clouds.
	\newblock {\em IEEE Transactions on Parallel and Distributed Systems}, pages
	1--16, 2019.
	
	\bibitem{Lu2017}
	D.~{Lu}, D.~{Huang}, A.~{Walenstein}, and D.~{Medhi}.
	\newblock A secure microservice framework for iot.
	\newblock In {\em 2017 IEEE Symposium on Service-Oriented System Engineering
		(SOSE)}, pages 9--18, April 2017.
	
	\bibitem{Pahl2018b}
	M.~{Pahl} and L.~{Donini}.
	\newblock Securing iot microservices with certificates.
	\newblock In {\em NOMS 2018 - 2018 IEEE/IFIP Network Operations and Management
		Symposium}, pages 1--5, April 2018.
	
	\bibitem{Nehme2019}
	A.~{Nehme}, V.~{Jesus}, K.~{Mahbub}, and A.~{Abdallah}.
	\newblock Securing microservices.
	\newblock {\em IT Professional}, 21(1):42--49, Jan 2019.
	
	\bibitem{Fetzer2016}
	C.~{Fetzer}.
	\newblock Building critical applications using microservices.
	\newblock {\em IEEE Security Privacy}, 14(6):86--89, Nov 2016.
	
	\bibitem{Nguyen2019}
	Quy Nguyen and Oras~F. Baker.
	\newblock Applying spring security framework and oauth2 to protect microservice
	architecture {API}.
	\newblock {\em {JSW}}, 14(6):257--264, 2019.
	
	\bibitem{He2017}
	Xiuyu He and Xudong Yang.
	\newblock Authentication and authorization of end user in microservice
	architecture.
	\newblock {\em Journal of Physics: Conference Series}, 910:012060, oct 2017.
	
	\bibitem{Baker2019}
	Oras Baker and Quy Nguyen.
	\newblock A novel approach to secure microservice architecture from owasp
	vulnerabilities.
	\newblock In {\em Proceedings of the 10th Annual CITRENZ Conference (2019)},
	pages 54--58, Nelson, NZ, October 2019. ITx New Zealand's Conference of IT.
	
	\bibitem{Salibindla2018}
	Jyothi Salibindla.
	\newblock Microservices api security.
	\newblock {\em International Journal of Engineering Research \& Technology},
	7(1):277--281, 2018.
	
	\bibitem{Jander2019}
	Kai Jander, Lars Braubach, and Alexander Pokahr.
	\newblock Practical defense-in-depth solution for microservice systems.
	\newblock {\em Journal of Ubiquitous Systems and Pervasive Networks}, 11(1):17
	-- 25, 2019.
	
	\bibitem{Torkura2018a}
	Kennedy~A Torkura, Muhammad~IH Sukmana, Feng Cheng, and Christoph Meinel.
	\newblock Cavas: Neutralizing application and container security
	vulnerabilities in the cloud native era.
	\newblock In {\em International Conference on Security and Privacy in
		Communication Systems}, pages 471--490. Springer, 2018.
	
	\bibitem{Chen2019}
	Jiyu Chen, Heqing Huang, and Hao Chen.
	\newblock Informer: Irregular traffic detection for containerized microservices
	rpc in the real world.
	\newblock In {\em Proceedings of the 4th ACM/IEEE Symposium on Edge Computing},
	SEC '19, pages 389--394, New York, NY, USA, 2019. ACM.
	
	\bibitem{Torkura2017}
	Kennedy~A. Torkura, Muhammad~I.H. Sukmana, and Christoph Meinel.
	\newblock Integrating continuous security assessments in microservices and
	cloud native applications.
	\newblock In {\em Proceedings of the10th International Conference on Utility
		and Cloud Computing}, UCC '17, pages 171--180, New York, NY, USA, 2017. ACM.
	
	\bibitem{Akkermans2018}
	Sven Akkermans, Bruno Crispo, Wouter Joosen, and Danny Hughes.
	\newblock Polyglot cerberos: Resource security, interoperability and
	multi-tenancy for iot services on a multilingual platform.
	\newblock In {\em Proceedings of the 15th EAI International Conference on
		Mobile and Ubiquitous Systems: Computing, Networking and Services},
	MobiQuitous '18, pages 59--68, New York, NY, USA, 2018. ACM.
	
	\bibitem{Beekman2017}
	Jethro~G. Beekman and Donald~E. Porter.
	\newblock Challenges for scaling applications across enclaves.
	\newblock In {\em Proceedings of the 2Nd Workshop on System Software for
		Trusted Execution}, SysTEX'17, pages 8:1--8:2, New York, NY, USA, 2017. ACM.
	
	\bibitem{Guija2018}
	Daniel Guija and Muhammad~Shuaib Siddiqui.
	\newblock Identity and access control for micro-services based 5g nfv
	platforms.
	\newblock In {\em Proceedings of the 13th International Conference on
		Availability, Reliability and Security}, ARES 2018, pages 46:1--46:10, New
	York, NY, USA, 2018. ACM.
	
	\bibitem{Li2019}
	Xing Li, Yan Chen, and Zhiqiang Lin.
	\newblock Towards automated inter-service authorization for microservice
	applications.
	\newblock In {\em Proceedings of the ACM SIGCOMM 2019 Conference Posters and
		Demos}, SIGCOMM Posters and Demos '19, pages 3--5, New York, NY, USA, 2019.
	ACM.
	
	\bibitem{Marquez2019}
	Gast\'{o}n M\'{a}rquez and Hern\'{a}n Astudillo.
	\newblock Identifying availability tactics to support security architectural
	design of microservice-based systems.
	\newblock In {\em Proceedings of the 13th European Conference on Software
		Architecture - Volume 2}, ECSA '19, pages 123--129, New York, NY, USA, 2019.
	ACM.
	
	\bibitem{Ibrahim2019}
	Amjad Ibrahim, Stevica Bozhinoski, and Alexander Pretschner.
	\newblock Attack graph generation for microservice architecture.
	\newblock In {\em Proceedings of the 34th ACM/SIGAPP Symposium on Applied
		Computing}, SAC '19, pages 1235--1242, New York, NY, USA, 2019. ACM.
	
	\bibitem{Stallenberg2019}
	Dimitri~Michel Stallenberg and Annibale Panichella.
	\newblock Jcomix: A search-based tool to detect xml injection vulnerabilities
	in web applications.
	\newblock In {\em Proceedings of the 2019 27th ACM Joint Meeting on European
		Software Engineering Conference and Symposium on the Foundations of Software
		Engineering}, ESEC/FSE 2019, pages 1090--1094, New York, NY, USA, 2019. ACM.
	
	\bibitem{Kramer2019}
	Michel Kr{\"a}mer, Sven Frese, and Arjan Kuijper.
	\newblock Implementing secure applications in smart city clouds using
	microservices.
	\newblock {\em Future Generation Computer Systems}, 99:308 -- 320, 2019.
	
	\bibitem{Jander2018}
	Kai Jander, Lars Braubach, and Alexander Pokahr.
	\newblock Defense-in-depth and role authentication for microservice systems.
	\newblock {\em Procedia Computer Science}, 130:456 -- 463, 2018.
	\newblock The 9th International Conference on Ambient Systems, Networks and
	Technologies (ANT 2018) / The 8th International Conference on Sustainable
	Energy Information Technology (SEIT-2018) / Affiliated Workshops.
	
	\bibitem{Abidi2019}
	Sarra Abidi, Mehrez Essafi, Chirine~Ghedira Guegan, Myriam Fakhri, Hamad Witti,
	and Henda Hjjami~Ben Ghezala.
	\newblock A web service security governance approach based on dedicated
	micro-services.
	\newblock {\em Procedia Computer Science}, 159:372 -- 386, 2019.
	\newblock Knowledge-Based and Intelligent Information \& Engineering Systems:
	Proceedings of the 23rd International Conference KES2019.
	
	\bibitem{Elsayed2019}
	Marwa Elsayed and Mohammad Zulkernine.
	\newblock Offering security diagnosis as a service for cloud saas applications.
	\newblock {\em Journal of Information Security and Applications}, 44:32 -- 48,
	2019.
	
	\bibitem{IBM2016}
	IBM.
	\newblock An integrated approach to insider threat protection, 2016.
	
	\bibitem{Kindervag2012}
	John Kindervag, Stephanie Balaouras, and Kelley Mak.
	\newblock Build security into your network's dna: The zero trust network
	architecture.
	\newblock Technical report, Forrester Research, November 2012.
	
	\bibitem{Zhuang2014}
	Rui Zhuang, Scott~A. DeLoach, and Xinming Ou.
	\newblock Towards a theory of moving target defense.
	\newblock In {\em Proceedings of the First ACM Workshop on Moving Target
		Defense}, MTD '14, pages 31--40, New York, NY, USA, 2014. Association for
	Computing Machinery.
	
	\bibitem{Mavroudis2017}
	Vasilios Mavroudis, Andrea Cerulli, Petr Svenda, Dan Cvrcek, Dusan Klinec, and
	George Danezis.
	\newblock A touch of evil: High-assurance cryptographic hardware from untrusted
	components.
	\newblock In {\em Proceedings of the 2017 ACM SIGSAC Conference on Computer and
		Communications Security}, CCS '17, pages 1583--1600, New York, NY, USA, 2017.
	Association for Computing Machinery.
	
	\bibitem{Merkel2014}
	Dirk Merkel.
	\newblock Docker: lightweight linux containers for consistent development and
	deployment.
	\newblock {\em Linux journal}, 2014(239):2, 2014.
	
\end{thebibliography}
\end{document}